\documentclass[journal,article,submit,pdftex,moreauthors]{Definitions/mdpi} 

\firstpage{1} 
\pubvolume{1}
\issuenum{1}
\articlenumber{0}
\pubyear{2024}
\copyrightyear{2024}
\datereceived{ } 
\daterevised{ } 
\dateaccepted{ } 
\datepublished{ } 
\hreflink{https://doi.org/} 
\pdfoutput=1 

\Title{Magnetar QPOs and neutron star crust elasticity}

\TitleCitation{Magnetar QPOs and neutron star crust elasticity}

\Author{Hajime Sotani $^{1,2,3}$\orcidA{}}

\AuthorNames{Hajime Sotani}

\AuthorCitation{Sotani, H.}

\address{%
$^{1}$ \quad Astrophysical Big Bang Laboratory, RIKEN, Saitama 351-0198, Japan\\
$^{2}$ \quad Interdisciplinary Theoretical \& Mathematical Science Program (iTHEMS), RIKEN, Saitama 351-0198, Japan\\
$^{3}$ \quad Theoretical Astrophysics, IAAT, University of T\"{u}bingen, 72076 T\"{u}bingen, Germany}

\corres{Correspondence: sotani@yukawa.kyoto-u.ac.jp}

\abstract{The crust region is a tiny fraction of neutron stars, but it has a variety of physical properties and plays an important role in astronomical observations. One of the properties characterizing the crust is the elasticity. In this review, with the approach of asteroseismology, we systematically examine neutron star oscillations excited by crust elasticity, adopting the Cowling approximation. In particular, by identifying the quasi-periodic oscillations observed in magnetar flares with the torsional oscillations, we make a constraint on the nuclear saturation parameters. In addition, we also discuss how the shear and interface modes depend on the neutron star properties. Once one detects an additional signal associated with neutron star oscillations, one can get a more severe constraint on the saturation parameters and/or neutron star properties, which must be a qualitatively different constraint obtained from terrestrial experiments, and help us to complementarily understand astrophysics and nuclear physics.}

\keyword{Neutron stars; Equation of state; Asteroseismology
} 

\begin{document}



\section{Introduction}

Neutron stars produced through the core-collapse supernova at the last moment of the massive star's life can become in extreme states, which is quite difficult to realize on Earth. For example, their density easily exceeds the standard nuclear density, $\sim 0.16$ fm$^{-3}$, and the gravitational and magnetic fields around/inside the star become much stronger than those observed in our solar system \cite{Shapiro83}. Therefore, physics in such extreme states may inversely be revealed through the observations of neutron stars and/or their phenomena.

The stellar structure depends on the equation of state (EOS) for neutron star matter, which is under beta-equilibrium and charge neutrality. However, the determination of EOS for neutron star matter from the experiments is still difficult. This is because the density of nuclei is around the saturation density almost independently of nuclear species, due to the nature of nuclear saturation properties. Namely, the experimental data concentrates around/below the saturation density, while the neutron star properties are mainly determined from the EOS for a higher-density region. Anyway, the saturation parameters characterizing the EOS in a lower-density region are gradually constrained through terrestrial experiments, e.g., \cite{SKC06,Tsang12,KM13,SSM14,Vinas14,New14,Tews17,OHKT17,Li19}, which are strongly associated with a low-mass (or low central density) neutron star properties.

Meanwhile, it is better to note that experiments cannot directly measure the saturation parameters, where one has to estimate them from a kind of empirical relations (or strong correlations) between the saturation parameters and experimental data, suggested by theoretical studies, e.g., \cite{Blaizot80,Piekarewicz07,GC18,RM11,RM13}. Thus, even if the experimental accuracy would be improved well, because of the theoretical uncertainties in these relations, it is not always true that the constraints on the saturation parameters become better \cite{Erl13,KS22,Naito22}. In practice, the estimation of the density-dependence of the nuclear symmetry energy, the so-called slope parameter $L$ (see Eq. (\ref{eq:E/A}) for the definition), from the parity asymmetry of the polarized electron scattering cross section of ${}^{208}$Pb strongly depends on the model \cite{PREXII,RRN21}. This means that the theoretical study must be vitally significant for deriving a technique to directly estimate the saturation parameters from the experimental data, although this is out of scope in this review.

On the other hand, astronomical observations must be important for constraining the EOS in a higher-density region. For example, the discovery of the massive neutron stars, whose mass is $\sim 2M_\odot$,  excluded the soft EOSs, with which the maximum mass does not reach the observed mass \cite{Demorest10,Antoniadis13,Cromartie20,Fonseca21}. The observation of gravitational waves from GW170817 \cite{GW170817} could restrict the dimensionless tidal deformability of neutron stars, which leads to the constraint that the $1.4M_\odot$ neutron star radius should be less than 13.6 km \cite{Annala18}. In addition, through the observation of the pulsar light curve, one may constrain the neutron star properties, especially the stellar compactness, $M/R$, with stellar mass $M$ and radius $R$, e.g., \cite{PF83,LL95,PG03,PO14,SM18,Sotani20}. This is because the trajectory of the photon radiating from the neutron star's surface can bend due to the strong gravitational field induced by the neutron star, which is a relativistic effect. In practice, the Neutron star Interior Composition Explorer (NICER) is now operating on an International Space Station, which successfully constrained the neutron star mass and radius for PSR J0030-0451 \cite{NICER19a,NICER19b} and for PSR J0740-6620 \cite{NICER21a,NICER21b}.

\begin{figure}[t]
\begin{center}
    \includegraphics[width=10 cm]{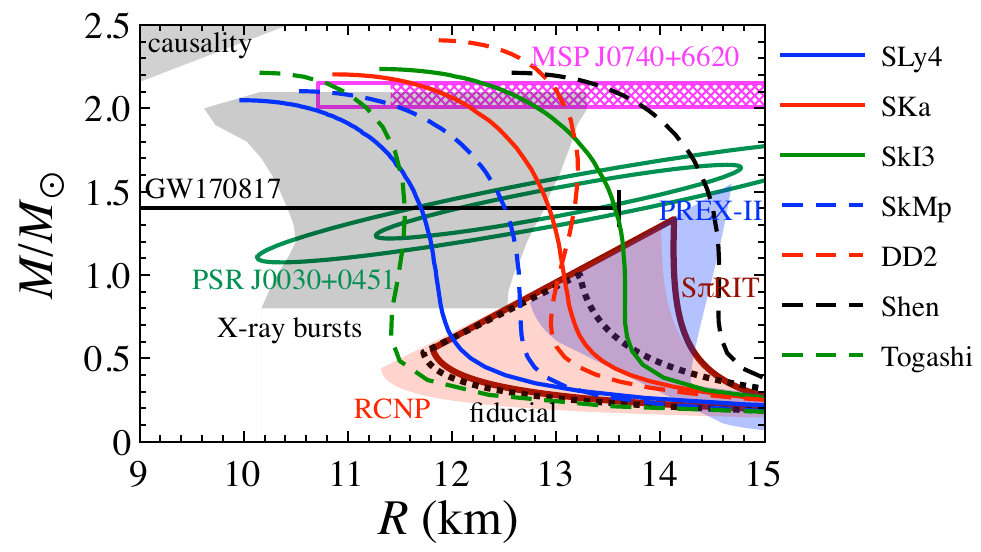}
\end{center}
\caption{Several constraints on the neutron star mass and radius obtained from the astronomical observations and terrestrial experiments. MSP J0740+6620 is one of the massive neutron stars discovered up to now, whose mass is $M/M_\odot = 2.08 \pm 0.07$ \cite{Fonseca21}, while the radius (and mass) of this object are also constrained via NICER as well as PSR J0030-0451. The tidal deformability restricted from the GW170817 event tells us the upper limit of the radius of $1.4M_\odot$ neutron star, i.e., $R_{1.4}\le 13.6$ km \cite{Annala18}. The observation of the X-ray bursts from neutron stars also gives us the constraint on the neutron star mass and radius, although they depend on the theoretical model \cite{SLB13}. In addition, one can theoretically exclude the top-left shaded region due to the causality \cite{L12}. On the other hand, the terrestrial experiments, e.g., by PREX-II \cite{PREXII}, by S$\pi$RIT \cite{SPiRIT}, and at RCNP \cite{RCNP}, give us the constraint on the stellar models constructed with a lower central density (bottom-right region), adopting the mass formula for low-mass neutron stars \cite{SIOO14}, where the constraints expected with the fiducial value of the nuclear saturation parameters, i.e., $L=60\pm20$ MeV \cite{Tsang12,New14,Li19} and $K_0=240\pm 20$ MeV \cite{SKC06}, is also shown (see Eq. (\ref{eq:E/A}) for the definition of $L$ and $K_0$). Furthermore, for reference, the mass and radius relations constructed with several EOSs, i.e., SLy4 \cite{Chabanat98,SLy4}, SKa \cite{SKa}, SkI3 \cite{SkI3}, SkMp \cite{SkMp}, DD2 \cite{DD2}, Shen \cite{Shen}, and Togashi \cite{Togashi}, are plotted.
Taken from \cite{SNN22}.\label{fig1}}
\end{figure}

In such a way, the neutron star mass and radius would be constrained more and more in a higher density region with astronomical observations and in a lower density region with the terrestrial experiments, e.g., \cite{SNN22,SO22,SN23} (also see Fig.~\ref{fig1}). Then, as these constraints from observations and experiments become more and more severe, one may eventually constrain the EOS for neutron star matter.

In addition to the observations (or estimations) of the neutron star mass and radius, the oscillation signals (and also the gravitational waves) from neutron stars, if observed, are another important information to extract the neutron star properties. Since the objects have their own specific oscillation frequencies, one may know the interior information via the observation of their frequencies as an inverse problem. This technique is known as asteroseismology, which is similar to seismology on Earth and helioseismology on the Sun. In fact, the neutron star has a lot of oscillation modes, where each mode can be excited due to the corresponding physics \cite{KS99}. 
For example, it is well-known that the frequency of fundamental oscillations in a neutron star is strongly associated with the square root of the stellar average density, because the fundamental oscillations are acoustic oscillations.
The gravitational waves from the neutron stars may be suitable astronomical information to adopt asteroseismology, but still, they have never been observed from an isolated neutron star. Even so, it has already been suggested that neutron star mass, radius, and EOS by observing the gravitational waves from the (cold) neutron stars, e.g., \cite{AK96,AK98,STM01,SKH04,SYMT11,PA12,DGKK13,KHA15,Sotani21,SK21,SD22}. Moreover, this technique is also adopted for extracting the information of protoneutron stars from the supernova gravitational waves, e.g., \cite{ST16,SKTK17,MRBV18,TCPOF19,SS19,ST20,STT21}.

Instead of the gravitational waves, the quasi-periodic oscillations (QPOs) discovered in the afterglow following the magnetar giant flares \cite{I05,SW05,SW06,H14a,HHWG14,MCS19,C21}, are also valuable information for adopting asteroseismology, where magnetars are strongly magnetized neutron stars. The observed QPO frequencies are in the range of tens of Hz up to kHz (see Sec.~\ref{sec:2} for details). 
Although the emission mechanism of the flare activity in magnetar is still under debate, the observed QPOs are considered to be strongly associated with neutron star oscillations. Taking into account the dynamical time of neutron stars, which is around 0.1 msec, it may be more difficult to theoretically explain the lower frequencies among the observed frequencies. The possible candidates may be either crustal torsional oscillations, magnetic oscillations, or magneto-elastic oscillations, if one assumes that the observed QPO frequencies come from the neutron star oscillations. However, the magnetic oscillations strongly depend on the magnetic field strength and its geometry inside the star, e.g., \cite{SKS07,G13,G18}, while their understanding is quite poor. Thus, here we discuss the crustal properties by identifying the observed QPO frequencies with the crustal torsional oscillations, neglecting magnetic effects (see Sec.~\ref{sec:2} for the magnetic effects on the crustal torsional oscillations). In this way, the crustal properties have been constrained indeed, e.g., \cite{SW09,G11,SNIO12,SNIO13a,SNIO13b,DSB14,SIO19}.

Through the approach with asteroseismology, it is shown that the observed QPO frequencies can be identified with the neutron star oscillations. On the other hand, there are still open issues in the magnetar QPOs. For instance, all the observed QPOs are not simultaneously excited after the magnetar flares, where some of the QPOs are excited first, then others. This phenomenon cannot be explained within the linear analysis. To understand the time dependence of the QPO frequency excitation, one may have to discuss beyond a linear analysis, such as a nonlinear coupling. Moreover, to discuss the crustal oscillations, one has to estimate the crustal elasticity, which is characterized by shear modulus. The details are shown in Sec.~\ref{sec:4}.  

The oscillation frequencies in a spherically symmetric neutron star model can be classified into two families with parity, i.e., axial and polar type oscillations, where the axial type oscillations can be excited independently of the polar type oscillations. The crustal torsional oscillations belong to the axial type oscillations.  Since the axial type oscillations do not involve the density variation, they are relatively easily excited but less important in the gravitational wave observations. On the other hand, polar-type oscillations involve density variation and stellar deformations, where various oscillation modes can exist.  

Owing to the crust elasticity, the shear and interface modes, which are polar type oscillations, are also excited as well as the torsional oscillations, even though they are not so discussed well compared to the torsional oscillations.
The interface modes, whose frequencies are less than 100 Hz, may be excited in binary neutron stars by resonating with the orbital frequency. In fact, 
the precursors have been observed just before the main flare activity of gamma-ray burst at a binary neutron star merger \cite{TRG10}, which may be a result of the resonant shattering of neutron star crusts induced by the binary orbital motion \cite{TRHPB12,T13}. If so, one could extract the neutron star properties by carefully observing the precursors and identifying them with neutron star oscillations \cite{Suvorov21,Neill21,Kuan22,Neill23}. Anyway, to extract the properties from the resonant shattering of neutron star crust, one has to examine the shear and interface modes systematically.

In this review, mainly focusing on the torsional oscillations excited in the neutron star, we extract the neutron star properties by identifying the observed QPO frequencies with the torsional oscillations. 
For this purpose, we simply assume the Cowling approximation in this review, i.e., we only consider the fluid oscillations with the (unperturbed) background metric. The perturbation equations are derived from the linearized energy-momentum conservation law. By imposing appropriate boundary conditions on the numerical boundary, the problem to solve becomes an eigenvalue problem. The neutron star crust is mainly composed of spherical nuclei, but the non-spherical nuclei may appear in the vicinity of the boundary of crust and core, depending on the nuclear parameters (or EOS). So, first, we start to discuss the observed QPO frequencies with the crustal torsional oscillations excited in the region only composed of spherical nuclei. Then, the discussion is extended to the more realistic situation with the non-spherical nuclei. Through these attempts, we successfully identify all the observed QPOs with the crustal torsional oscillations and finally derive the constraint on the slope parameter as $L\simeq 58-73$ MeV. Furthermore, we also discuss the identification of the high-frequency QPOs observed in GRB 200415A with the overtones of crustal torsional oscillations. Through this identification, we make a constraint on the neutron star mass and radius for GRB 200415A, adopting the fiducial values of nuclear saturation parameters.
In addition, we systematically examine the shear and interface modes for various neutron star models, with which we find the relation between the frequencies and the neutron star properties independently of the stiffness of EOS for core region.

This manuscript is organized as follows. In Sec. \ref{sec:2}, we shortly mention the observed QPO frequencies in magnetar giant flares, which will be considered as evidence of neutron star oscillations. In Sec. \ref{sec:3}, we show the equilibrium neutron star models and the EOS considered in this study. In Sec. \ref{sec:4}, we briefly mention the shear modulus inside the neutron star crust, which is an important property for considering the crustal oscillations. In Sec. \ref{sec:5}, we extract the crustal information in practice by identifying the observed QPOs with the crustal torsional oscillations. In Sec. \ref{sec:6}, we also show the behavior of the shear and interface modes, especially focusing on the association with the neutron star properties. Finally, we conclude in Sec. \ref{sec:7}. Unless otherwise mentioned, we adopt geometric units in the following, $c=G=1$, where $c$ and $G$ denote the speed of light and the gravitational constant, respectively.

\section{Magnetar QPOs and magneitc effects on the crustal oscillations}
\label{sec:2}

Compared to usual pulsars whose surface magnetic field strength is $\sim 10^{12}-10^{13}$ G, the existence of neutron stars with strong magnetic fields, such as $\sim 10^{14}-10^{15}$ G, is observationally known. This strongly magnetized neutron star is known as a magnetar. In a rotating neutron star, the magnetic stress induced by the stellar rotation gradually accumulates in the crust, which is usually supported by crustal elasticity. But, once the magnetic stress becomes significantly strong and the crust elasticity cannot support it, the crust eventually breaks out, and the magnetic energy is released through the flare activities~\cite{DT92,TD95}, where star quakes may also happen. This scenario could be the origin of the QPOs observed in the afterglow following the magnetar giant flares.

In practice, the QPOs have been discovered in the giant flares observed in soft-gamma repeaters (SGRs). 
At least, there are three events detected up to now, in which the QPOs are found, i.e., SGR 0526-66 in 1979~\cite{Mazets79,Barat83}, SGR 1900+14 in 1998~\cite{Hurley99}, and SGR 1806-20 in 2004~\cite{Terasawa05,Palmer05}. 
In particular, in the events on SGR 1900+14 and SGR 1806-20, several QPOs are found~\cite{I05,SW05,SW06}, i.e., 28, 54, 84, and 155 Hz in SGR 1900+14, and 18, 26, 29 (or 30), 92.5, 150, 626.5, and 1837 Hz in SGR 1806-20. By a Bayesian analysis, the existence of additional QPOs was also reported in SGR 1806-20 \cite{MCS19}. Moreover, even without giant flares, the QPOs have been found from less energetic recurrent bursts, i.e., 93 and 127 Hz in SGR J1550-5418 \cite{H14a}, and 57 Hz in SGR 1806-20 \cite{HHWG14}. In addition to the QPOs observed in the SGRs, several high QPO frequencies have been found in GRB 200415A, i.e., 836, 1444, 2132, and 4250 Hz \cite{C21}.

Assuming that these QPOs come from neutron star oscillations, one may identify the higher QPO frequencies with polar-type oscillations of neutron stars, e.g., Fig.~\ref{fig:QPO-is}, but the lower frequencies are more difficult to identify with neutron star oscillations. Maybe the possible candidates are only the crustal torsional oscillations or magnetic oscillations (or magneto-elastic oscillations). Meanwhile, the magnetic oscillations definitely depend on the geometry and strength of magnetic fields, which have not been understood well yet. In addition, it is pointed out that crustal torsional oscillations can be excited in the vicinity of the neutron star surface, while magnetic oscillations are confined only to the core region, if magnetic fields are not so strong. 
This is because the shear velocity is higher than the Alfv\'{e}n velocity, $v_A\equiv B/\sqrt{4\pi \rho}$, with the magnetic field strength, $B$, and density, $\rho$, at the basis of the crust, which leads to that the oscillations are controlled by shear properties. In general, 
the critical field strength, above which the magnetic oscillations become dominant, depends on the EOS (or crust properties), which is considered around a few times $10^{14}$ G \cite{SKS07,G18,SKS23}. So, to avoid the uncertainty in the magnetic geometry and profile, assuming that the field strength is not so strong, we focus on only the crustal oscillations without magnetic effects in this review.

As an advantage without the magnetic effect, the crustal torsional oscillations are completely confined inside the crust, since the axial-type oscillations do not involve radial variations. That is, one can avoid the large uncertainties in the EOSs of the core region, and directly discuss the crust properties by comparing the observed frequencies to the crustal torsional oscillations. On the other hand, if one takes into account the magnetic effects, even torsional oscillations can be coupled with the core region through the magnetic fields, where oscillations easily damped out via magnetic coupling if the magnetic fields are relatively larger \cite{Gab11,G12}. We note that one has to consider the polar-type oscillations in the whole region of a neutron star, even if the magnetic effects are neglected, because the polar-type oscillations involve radial variations.  

We only focus on the neutron star oscillations associated with the crust elasticity in this review, while we briefly mention the magnetic oscillations in the neutron star here. Since the existence of magnetic fields inside the star breaks a spherically symmetric configuration, one cannot examine the magnetic oscillations (and also magneto-elastic oscillations) in the same way as the case on the spherically symmetric objects (as mentioned in the following sections). Assuming a specific distribution of magnetic fields, one may study them by decomposing the perturbative variables with spherical harmonic function, where $\ell$-th order perturbation equations are generally coupled with $\ell$-th order equations of other values \cite{SKS07,Lee08}. Or, one may solve a time evolution of two-dimensional perturbation equations and determine the frequencies via FFT. 

With the later approach, it has been shown that the axial-type magnetic oscillations are continuous spectrum instead of the discrete one \cite{SKS08,CBK09,CSF09}. This is because the frequencies of magnetic oscillations are characterized by the propagation time determined with the magnetic field length inside the star and Alfv\'{e}n velocity, while the magnetic field length inside the star continuously changes from the axis to the position related to the closed field line close to the equatorial plane \cite{CBK09}. This phenomenon may also be understood with a toy model suggested in \cite{Levin06,Levin07}. In any case, taking into account the coupling with the crust, i.e., magneto-elastic oscillations, one may identify the observed QPO frequencies \cite{CK11,CK12,G12,G13}. Furthermore, one may constrain the magnetic geometry from the QPO frequencies and estimated field strength of magnetic fields, i.e., the case that the magnetic fields are confined in the crust due to type I superconductor in the core region can be excluded \cite{SCK08}. 

Compared to the extensive studies about axial-type magnetic oscillations, the study of polar-type magnetic oscillations is relatively poor. Even so, unlike the axial-type oscillations, the polar-type magnetic oscillations seem to be discrete spectrum \cite{SK09}, and they definitely play an important role in the gravitational wave radiations \cite{LvH11,KK11}.

\section{Crustal equilibrium}
\label{sec:3}

\begin{figure}[t]
\begin{center}
    \includegraphics[width=6 cm]{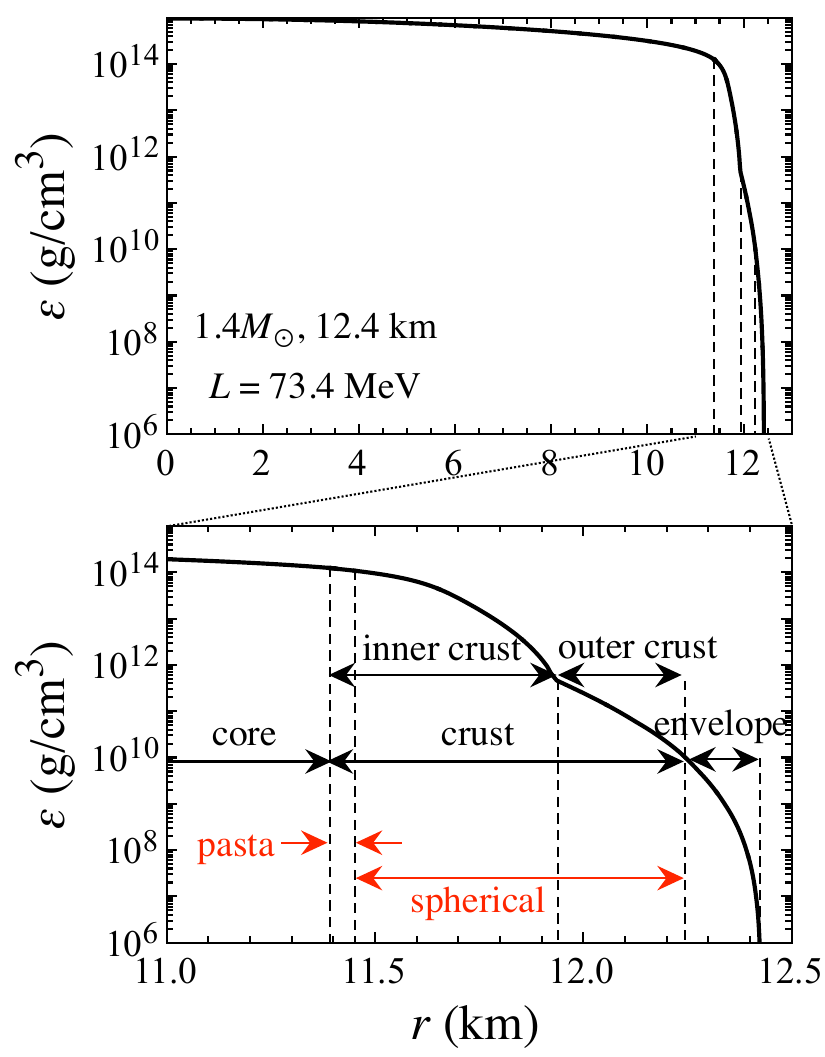}
\end{center}
\caption{Example of the energy density profile inside the neutron star with $1.4M_\odot$ and $12.4$ km constructed with a specific EOS with $L=73.4$ MeV and $K_0=230$ MeV, where $L$ and $K_0$ are the nuclear saturation parameters given by Eq. (\ref{eq:E/A}). The bottom panel is an enlarged view of the above panel. 
Taken from \cite{Sotani23}.\label{fig2}}
\end{figure}

Although the details of a neutron star structure depend on the EOS for neutron star matter, its conceptual structure is generally accepted. Under the thin envelope of the atmosphere and/or ocean, the matter forms a Coulomb lattice, which behaves as a solid. This region corresponds to the neutron star crust. Most of the crust region is composed of spherical nuclei with a body-centered cubic (bcc) lattice, while the shape of stable nuclei changes into cylindrical, slablike, cylindrical-hole, and spherical-hole, and eventually becomes uniform matter, as density increases \cite{O93,PR95}. The region composed of non-spherical nuclei is the so-called pasta phase, which behaves as a liquid crystal. Meanwhile, the region whose density is higher than the transition density from the non-uniform to the uniform matter is the neutron star core. This transition density depends on the neutron star EOS, which becomes $\sim (1/3-1/2)$ of the nuclear saturation density. The transition density where the nuclear shape changes inside the crust also depends on the EOS, especially on the density dependence of the symmetry energy, $L$, in Eq. (\ref{eq:E/A}) \cite{OI03,OI07,Oya23}. In addition, the crust thickness is at most $10\%$ of the stellar radius, which depends on the stellar compactness, $M/R$, and the nuclear saturation parameters \cite{PR95,SIO17b} (see Fig.~\ref{fig2} for an example of the concrete stellar model). In general, the ratio of crust thickness to the stellar radius decreases as the stellar compactness and $L$ increases. In particular, for $L\gtrsim 100$ MeV, the pasta phase disappears, where the crust is composed of only the spherical nuclei, and directly transitions into uniform matter \cite{OI03,OI07,Oya23} (also see Table~\ref{tab:SH-density}). 

The EOS for neutron star matter generally depends on the adopted nuclear interaction, nuclear models, and the compositions. But for any EOSs, the bulk energy per baryon for the zero-temperature nuclear matter can be expanded in the vicinity of the saturation density, $n_0\simeq 0.16$ fm$^{-3}$, of symmetric nuclear matter as a function of the baryon number density, $n_b$, and neutron excess, $\alpha$, via \cite{L81}
\begin{equation}
  \frac{E}{A} = w_0 + \frac{K_0}{2}u^2 + {\cal O}(u^3) 
     +  \left[S_0 + Lu + {\cal O}(u^2)\right]\alpha^2 + {\cal O}(\alpha^3), 
\label{eq:E/A}
\end{equation}
where $u\equiv(n_b-n_0)/(3n_0)$, while $n_b$ and $\alpha$ are given by $n_b=n_n+n_p$ and $\alpha=(n_n-n_p)/n_b$, using the neutron and proton number density, $n_n$ and $n_p$. We note that the matter with $\alpha=0$ corresponds to a symmetric nuclear matter, i.e. $n_n=n_p$, while that with $\alpha=1$ is a pure neutron matter, i.e., $n_b=n_n$. The coefficients in this expansion are the nuclear saturation parameters, which characterize each EOS, and the term inside the brackets in front of $\alpha^2$ corresponds to the nuclear symmetry energy. 
In particular, $w_0$ and $K_0$ are the binding energy and incompressibility of symmetric nuclear matter at $n_b=n_0$, while $S_0$ is the symmetry energy at $n_b=n_0$.

Owing to the nature of the nuclear saturation properties, $n_0$, $w_0$, and $S_0$ are well determined via experiments, i.e., $n_0\approx 0.15-0.16$ fm$^{-3}$, $w_0\approx -15.8$ MeV \cite{OHKT17}, and $S_0\approx 31.6\pm 2.7$ MeV \cite{Li19}. In contrast, $K_0$ and $L$ are more difficult to determine experimentally, because they are a density derivative at $n_b=n_0$, i.e., one needs to know the nuclear information in a somewhat wide density range around the saturation point. Nevertheless, the constraints on these two parameters are improved, where their fiducial values are $K_0=230\pm 40$ MeV \cite{SKC06,KM13} and $L\simeq 60\pm 20$ MeV \cite{Tsang12,New14,Li19}. We note that the constraint on $L$ still seems to have large uncertainties, e.g., $L=106\pm 37$ MeV with PREX-II done by the Thomas Jefferson National Accelerator Facility in Virginia \cite{PREXII}, or $42\le L \le 117$ MeV with S$\pi$RIT by Radioactive Isotope Beam Factory at RIKEN in Japan \cite{SPiRIT}.

Anyway, in this review, to systematically discuss the dependence of the crustal oscillations on the nuclear saturation parameters, we especially adopt a phenomenological EOS proposed by Oyamatsu and Iida \cite{OI03,OI07} (hereafter referred to as OI-EOS). For given values of $K_0$ and $L$, the OI-EOS family is constructed by optimizing the values of $n_0$, $w_0$, and $S_0$ to reproduce the experimental data for masses and charge radii of stable nuclei in such a way that the bulk energy reduces to Eq. (\ref{eq:E/A}) in the limit of $n_b\to n_0$ and $\alpha\to 0$, adopting the extended Thomas-Fermi theory. The EOS parameters adopted in this review are listed in Table \ref{tab:SH-density}. We note that some of the EOS parameter sets are out of the range of fiducial values for $K_0$ and $L$, but we adopt such a wide range to systematically discuss the dependence on the saturation parameters.

\begin{table}
\centering
\caption{
The EOS parameters for the OI-EOS family adopted in this review, where 
$y$ is defined as $y\equiv-K_0S_0/(3n_0L)$.
SP-C, C-S, S-CH, CH-SH, and SH-U denote the transition densities for 
each EOS parameter set.  The asterisk at the value of $K_0$ denotes the EOS model with 
which some pasta phases are not predicted to appear, i.e., the values 
with $*1$, $*2$, and $*3$ denote the transition densities from 
cylindrical nuclei to uniform matter, from cylindrical-hole nuclei to uniform 
matter, and from spherical nuclei to uniform matter, respectively.
}
\scalebox{0.75}{
\begin{tabular}{ccccccccc}
\hline\hline
  $K_0$ (MeV) & $-y$ (MeV fm$^3$) & $L$ (MeV) & SP-C (fm$^{-3}$) & C-S (fm$^{-3}$) & S-CH (fm$^{-3}$) & CH-SH (fm$^{-3}$) & SH-U (fm$^{-3}$)  \\
\hline
  180 & 1800 & 5.7   & 0.06000 & 0.08665 & 0.12039 & 0.12925   & 0.13489     \\  
  180 & 600 & 17.5 & 0.05849 & 0.07986 & 0.09811 & 0.10206  &  0.10321     \\  
  180 & 350 & 31.0 & 0.05887 & 0.07629 & 0.08739 & 0.09000   & 0.09068     \\  
  180 & 220 & 52.2 & 0.06000 & 0.07186 & 0.07733 & 0.07885   & 0.07899     \\  
\hline
  230 & 1800 & 7.6   & 0.05816 & 0.08355 & 0.11440 & 0.12364   & 0.12736     \\  
  230 & 600 & 23.7 & 0.05957 & 0.07997 & 0.09515 & 0.09817  &  0.09866     \\  
  230 & 350 & 42.6 & 0.06238 & 0.07671 & 0.08411 & 0.08604   & 0.08637     \\  
  230 & 220 & 73.4 & 0.06421 & 0.07099 & 0.07284 & 0.07344   & 0.07345     \\  
\hline
  280 & 1800 & 9.6  & 0.05747 & 0.08224 & 0.11106 & 0.11793  &  0.12286   \\ 
  280 & 600 & 30.1 & 0.06218 & 0.08108 & 0.09371 & 0.09577 &  0.09623    \\ 
  280 & 350 & 54.9 & 0.06638 & 0.07743 & 0.08187 & 0.08314 &   0.08331   \\ 
  $^*$280 & 220 & 97.5 & 0.06678 & --- & ---  & ---  &  0.06887$^{*1}$  \\ 
\hline
  360 & 1800 & 12.8 & 0.05777 & 0.08217 & 0.10892 & 0.11477   & 0.11812     \\  
  360 & 600 & 40.9 & 0.06743 & 0.08318 & 0.09197 & 0.09379  &  0.09414     \\  
  $^*$360 & 350 & 76.4   & 0.07239 & 0.07797 & 0.07890 & --- & 0.07918$^{*2}$     \\  
  $^*$360 & 220 & 146.1 & --- & --- & --- & ---   & 0.06680$^{*3}$     \\  
\hline\hline
\end{tabular}
}
\label{tab:SH-density}
\end{table}

As an equilibrium crust model (or a neutron star model), we simply consider a non-rotating, strain-free, and spherically symmetric neutron star model in this review. The metric describing such a stellar model is given by 
\begin{equation}
  ds^2 = -e^{2\Phi}dt^2 + e^{2\Lambda}dr^2 + r^2(d\theta^2 + \sin^2\theta d\phi^2), \label{eq:metric}
\end{equation}
where $\Phi$ and $\Lambda$ are the metric functions depending on only the radial coordinate, $r$. $\Lambda$ is directly associated with the mass function, $m(r)$, which corresponds to the enclosed gravitational mass inside the position $r$, i.e., $e^{-2\Lambda}=1-2m/r$. Adopting an appropriate EOS, e.g., the OI-EOS family in this review, one can construct the stellar model by integrating the Tolman-Oppenheimer-Volkoff (TOV) equation. In general, to construct a stellar model, one integrates the TOV equation outward from the center to the surface, assuming a central density, i.e., the neutron star models become one parameter family for each EOS. In fact, we construct the stellar model in this way for discussing the shear and interface modes in Sec. \ref{sec:6}, because these modes are polar-type oscillations, which are excited not only in the crust region but also inside the core. On the other hand, the torsional oscillations discussed in Sec. \ref{sec:5}, which are axial-type oscillations, are confined inside the crust (elastic) region. So, to remove the uncertainties of the core EOS, we construct only the crust equilibrium for discussing the torsional oscillations, where we integrate the TOV equations inward from the stellar surface to the base of the crust, assuming the stellar mass and radius as a boundary condition at the stellar surface.

\section{Shear modulus}
\label{sec:4}

The elasticity inside the crust can be characterized by the shear modulus, $\mu$. The shear modulus, $\mu_{sp}$, in a phase of spherical nuclei with a bcc lattice is approximately described in the limit of zero-temperature as
\begin{equation}
  \mu_{sp} = 0.1194\frac{n_i(Ze)^2}{a}, \label{eq:mu_sp}
\end{equation}
where $n_i$, $Z$, and $a$ are the ion number density, charge number of the ion, and Wigner-Seitz cell radius, i.e., $4\pi a^3/3 = 1/n_i$ \cite{S91}. This expression has been derived by averaging the overall direction, assuming that each nuclei is only a point-particle  \cite{OI90}. The shear modulus in the phase of spherical nuclei has been discussed more by including the phonon contribution \cite{Baiko11}, the electron-screening effect \cite{KP13,S14,SIO16,Tews17a}, the effect of polycrystal properties \cite{KP15}, and effects of finite size of atomic nuclei \cite{STT22}, which can modify the shear modulus more or less. But, in this review, we simply adopt $\mu_{sp}$ given by Eq. (\ref{eq:mu_sp}) as shear modulus in the phase of spherical nuclei.

Compared to the shear modulus in the phase of spherical nuclei, the discussion for the shear modulus in the pasta phases is very poor. Nevertheless, according to Ref. \cite{PP98}, the shear modulus, $\mu_{cy}$, in the phase of cylindrical nuclei can be estimated as
\begin{equation}
  \mu_{cy} = \frac{2}{3}E_{\rm Coul}\times 10^{2.1(w_2-0.3)}, \label{eq:mu_cy}
\end{equation}
where $E_{\rm Coul}$ and $w_2$ denote the Coulomb energy per volume of a Wigner-Seitz cell and the volume fraction of cylindrical nuclei, while the shear modulus, $\mu_{sl}$, in the phase of slablike nuclei is
\begin{equation}
  \mu_{sl} = 0, \label{eq:mu_sl}
\end{equation}
i.e., the matter composed of slablike nuclei behaves as a fluid within the linear response. This is because the deformation energy in slablike nuclei becomes of higher order with respect to the displacement \cite{PP98}. So, in this review we simply adopt Eqs. (\ref{eq:mu_cy}) and (\ref{eq:mu_sl}) as the shear modulus in the phases of cylindrical and slablike nuclei. But, on the other hand, it has also been suggested that the elastic constant in the poly-crystalline lasagna (which corresponds to slablike nuclei) may become a (nonzero) tiny value \cite{CSH18,PZK20}. If one considers the nonzero shear modulus inside the slablike nuclei, the results shown in this review may be changed.

Furthermore, the shear modulus in the phase of cylindrical-hole, $\mu_{ch}$, and spherical-hole nuclei, $\mu_{sh}$, may be considered in a similar way to that in the phase of cylindrical and spherical nuclei. This is because the liquid crystalline structures in the phase of cylindrical-hole and spherical-hole nuclei are basically the same as in the phase of cylindrical and spherical nuclei. Thus, we adopt Eqs. (\ref{eq:mu_cy}) and (\ref{eq:mu_sp}) for $\mu_{ch}$ and $\mu_{sh}$ by appropriately replacing the quantities in the formulae \cite{SIO17a,SIO19}. That is, $w_2$ in Eq. (\ref{eq:mu_cy}) is replaced by the volume fraction of gas of dripped neutron for $\mu_{ch}$, while $n_i$ and $Z$ in Eq. (\ref{eq:mu_sp}) are replaced by the number density of spherical-holes (bubbles) and the effective charge number, $Z_{sh}$, of spherical-hole for $\mu_{sh}$, where $Z_{sh}$ is estimated as $Z_{sh}=n_QV_{sh}$ with the effective charge number density of the spherical-hole, $n_Q$, and the volume of the spherical-hole, $V_{sh}$. $n_Q$ is given by the difference of the charge number density
inside the spherical-hole from that outside the spherical-hole, i.e., $n_Q = -n_e - (n_p - n_e) = -n_p$, using the proton number density outside the spherical-hole, $n_p$, and the number density of uniform electron gas, $n_e$.

\begin{figure}[t]
\begin{center}
\includegraphics[width=12 cm]{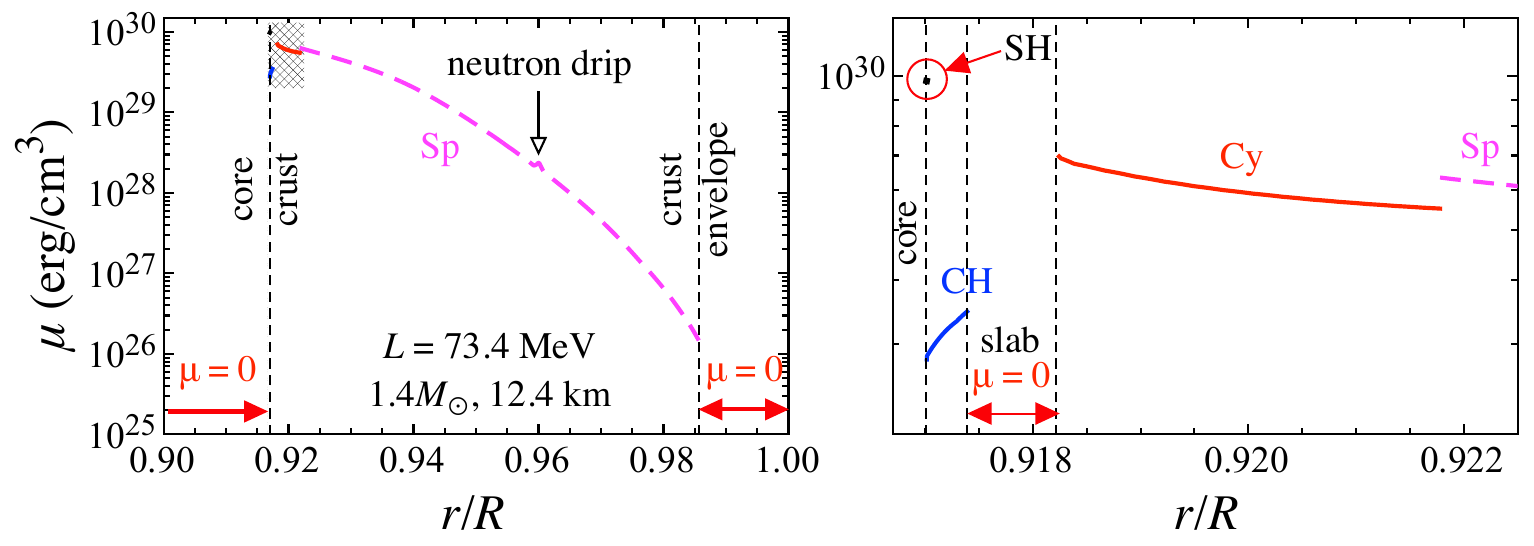}
\end{center}
\caption{Shear modulus inside the neutron star model given by Fig.~\ref{fig2}, where Sp, Cy, CH, and SH respectively denote the phases composed of spherical, cylindrical, cylindrical-hole, and spherical-hole nuclei. The right panel is an enlarged view of the shaded region in the left panel. Taken from \cite{Sotani23}. \label{fig3}}
\end{figure}

In Fig.~\ref{fig3}, we show an example of the distribution of shear modulus inside the crust for the stellar model shown in Fig.~\ref{fig2}, where the right panel is an enlarged view of the shaded region in the left panel. 

\section{Crustal torsional oscillations}
\label{sec:5}

On the crustal equilibrium models mentioned in the previous section, we make a linear perturbation analysis to determine the specific frequencies of crustal torsional oscillations. 
In this review, we examine the oscillation frequencies within the relativistic framework, while it has also been discussed within the Newtonian framework, e.g., \cite{HC80,MvHH88,S91}. To derive the perturbation equations, 
we simply adopt the Cowling approximation, where the metric perturbation is neglected during the fluid oscillations. 
We note that one can expect to accurately determine the frequencies of the axial-type oscillations even with the Cowling approximation, because the axial-type oscillations do not involve the density variations, i.e., the metric perturbations (corresponding to the perturbations of gravitational potential) should be too small.   
The perturbation equation can be derived from the linearized energy-momentum conservation law. In practice, introducing the $\phi$-component of the Lagrangian displacement, $\xi^\phi = {\cal Y}(r)e^{i\omega t}\partial_\theta P_\ell(\cos\theta)/\sin\theta$, where $P_\ell(\cos\theta)$ denotes the $\ell$-th order Legendre polynomial, the perturbation equation becomes \cite{ST83,SKS07}
\begin{equation}
  {\cal Y}'' 
  +\left[\left(\frac{4}{r} + \Phi' - \Lambda'\right)
  +\frac{\mu'}{\mu}\right]{\cal Y}' 
  +\left[\frac{\tilde{H}}{\mu}\omega^2e^{-2\Phi} - \frac{(\ell+2)(\ell-1)}{r^2}\right]e^{2\Lambda}{\cal Y}=0, \label{eq:6}
\end{equation}
where the dash denotes the derivative with $r$ and $\tilde{H}$ denotes the effective enthalpy contributing the oscillations, given by
\begin{equation}
  \tilde{H}
    = \left(1-\frac{N_s}{A}\right)H
  \label{eq:7}
\end{equation}
for the spherical or cylindrical nuclei \cite{SNIO13a,SIO18}, while 
\begin{equation}
  \tilde{H}
    =\frac{N_i+{\cal R}(A-N_i)}{A}H
  \label{eq:8}
\end{equation}
for the cylindrical-hole or spherical-hole nuclei \cite{SIO19}.

In the expression of the effective enthalpy, $H$ denotes the local enthalpy given by $H=\varepsilon + p$ with the energy density, $\varepsilon$, and pressure, $p$, while $A$ is the baryon number density inside a Wingner-Seitz cell, $N_s$ is the number of neutrons inside a Wingner-Seitz cell that do not comove with protons in the nuclei, $N_i$ is the number of neutrons inside a cylindrical-hole or spherical-hole, and ${\cal R}$ is a parameter characterizing a participant ratio in the oscillations, i.e., how much ratio of nucleons outside the cylindrical-hole or spherical-hole comove with it non-dissipatively. That is, all the nucleons inside a Wingner-Seitz cell contribute to the effective enthalpy with ${\cal R}=1$ (maximum enthalpy), while no nucleons outside the cylindrical-hole or spherical-hole do so with ${\cal R}=0$ (minimum enthalpy). We note that ${\cal R}$ in a phase of spherical-hole nuclei is predicted as $\sim 0.34-0.38$ at $n_b=0.08$ fm$^{-3}$ from the band calculations \cite{Chamel12}. Meanwhile, assuming that $N_s$ comes only from a part of the dripped neutrons inside a Wingner-Seitz cell, $N_d$, the parameter $N_s/N_d$ can control the fraction of neutrons contributing to the oscillations, i.e., $N_s/N_d=0$ corresponds to the situation that all the dripped neutrons comove with the protons (maximum enthalpy), while $N_s/N_d=1$ is that all the dripped neutrons behave as a superfluid and do not contribute to the oscillations (minimum enthalpy).

To determine the frequencies, one has to impose appropriate boundary conditions in the perturbation equation (\ref{eq:6}) at the basis of the crust (the boundary between the core and crust) and the surface of the crust (or the boundary between the crust and envelope). Both boundaries are essentially equivalent to the boundary between the elastic and fluid region, where we impose the condition that the traction force vanishes \cite{ST83,SKS07}, which reduces to ${\cal Y}'=0$. On the other hand, at the interface between the elastic regions with different shapes of nuclei, e.g., between the spherical and cylindrical nuclei, we impose a continuous traction condition, i.e., $\mu{\cal Y}'$ should be continuous at such an interface. Then, the problem to solve becomes an eigenvalue problem with respect to the eigenvalue, $\omega$. The resultant value of $\omega$ gives us the frequency, $f$, via $f=\omega/2\pi$ for each value of $\ell$. In this review, we use the notation of ${}_n t_\ell$ for expressing the eigenfrequencies of torsional oscillations with the angular index $\ell$ and the nodal number in the eigenfunction (although the notation is different from the usual one, such as ${}_\ell t_n$).

\subsection{Torsional oscillations excited in the spherical nuclei}
\label{sec:5a}

\begin{figure}[t]
\begin{center}
    \includegraphics[width=12 cm]{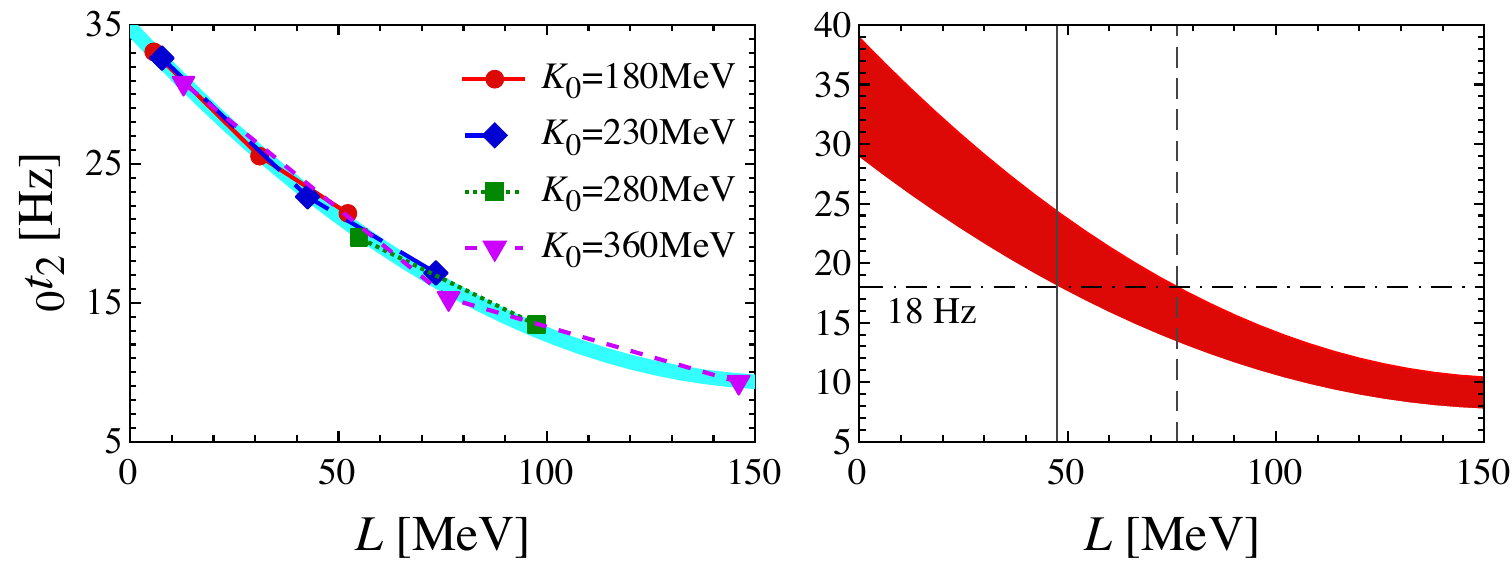}
\end{center}
\caption{ 
In the left panel, the fundamental frequencies of $\ell=2$ torsional oscillations for a neutron star model with $1.4M_\odot$ and 12 km constructed with several EOS parameter sets are shown as a function of $L$. The thick-solid line is the expected frequency using the fitting formula given by Eq. (\ref{eq:9}). In the right panel, the expected fundamental frequency of $\ell=2$ torsional oscillations as a function of $L$, assuming that the neutron star mass and radius are in the range of $1.4\le M/M_\odot \le 1.8$ and 10 km $\le R \le$ 14 km, where for reference the lowest QPO frequency observed in SGR 1806-20, i.e., 18 Hz, is also shown. The vertical solid and broken lines correspond to $L = 47.4$ and 76.2 MeV, respectively.
Taken from \cite{SNIO13b}.\label{fig4}}
\end{figure}

Even though the crust is not only composed of the spherical nuclei as mentioned above, to see the dependence of the frequencies of the torsional oscillations on the EOS parameters and stellar properties, first we consider the torsional oscillations excited in the elastic region composed of only the spherical nuclei. As a first step, we examine the frequencies without neutron superfluidity, i.e., $N_s/N_d=0$ (or $\tilde{H}=H$). As mentioned in Sec. \ref{sec:3}, to construct the equilibrium model, we have to select two parameters for EOS, i.e., $K_0$ and $L$, and two for stellar models, $M$ and $R$. In the left panel of Fig.~\ref{fig4}, we show the fundamental frequencies of $\ell = 2$ torsional oscillations for a neutron star model with $1.4M_\odot$ and 12 km, using the EOSs with some parameter sets listed in Table \ref{tab:SH-density} (see Ref. \cite{SNIO13b} for the concrete parameter sets adopted here), as a function of $L$. From this figure, one can observe that the frequencies are less sensitive to $K_0$. Thus, we will discuss the dependence of the fundamental frequency as a function of $L$ through the fitting formula by assuming the polynomial form as
\begin{equation}
  {}_0t_2 = c_2^{(0)} + c_2^{(1)} L_{100} + c_2^{(2)} L_{100}^2
  \label{eq:9}
\end{equation}
where $c_2^{(i)}$ with $i=0,1,2$ are the adjustable parameters depending on $M$ and $R$ (and $N_s/N_d$), while $L_{100}$ is the value of $L$ normalized by 100 MeV. Using this fitting formula, one can estimate the frequency of ${}_0t_2$ less than $\sim 5\%$ accuracy for a neutron star model with $1.4M_\odot$ and 12 km (Table 2 in \cite{SNIO13b}). In the left panel of Fig.~\ref{fig4}, we also show the expected frequencies with the fitting formula given by Eq. (\ref{eq:9}) by the thick-solid line.

One may be able to understand the dependence of ${}_0t_2$ on $L$ as follows. As the value of $L$ increases, the nuclear symmetry energy conversely decreases at the sub-nuclear density, considering the density dependence of the bulk energy given by Eq. (\ref{eq:E/A}). As a result, protons easily turn into neutrons, which leads to the situation that the charge number of nuclei gets smaller. This means that $\mu_{sp}$ becomes smaller from Eq. (\ref{eq:mu_sp}). On the other hand, the fundamental frequency of the $\ell$-th torsional oscillations, ${}_0t_\ell$, are estimated as ${}_0t_\ell\sim [\ell(\ell+1)]^{1/2}v_s/R$ with the shear velocity, $v_s\equiv (\mu/\tilde{H})^{1/2}$ \cite{HC80}. Thus, one can expect that ${}_0t_\ell$ decreases with $L$.

Now, assuming the neutron star mass and radius would be in the range of $1.4\le M/M_\odot \le 1.8$ and 10 km $\le R \le$ 14 km, which are reasonable assumptions as a neutron star model, the resultant fundamental frequency of the $\ell=2$ torsional oscillations is shifted, depending on the stellar models. In the right panel of Fig.~\ref{fig4}, we show the expected region of ${}_0t_2$ for such neutron star models as a function of $L$. We note that ${}_0t_2$ decreases with $R$ and $M$. So, the lower bound of the shaded region corresponds to the stellar model with $1.8M_\odot$ and $14$ km. On the other hand, the lowest QPO frequency observed in the SGR 1806-20, i.e., 18 Hz, is also shown in the right panel of Fig.~\ref{fig4}. Since the $\ell=2$ fundamental torsional oscillations are theoretically the lowest frequency among a lot of eigenfrequencies of the torsional oscillations, ${}_0t_2$ should be equal to or even lower than the observed lowest QPO frequency on the assumption that the magnetar QPOs come from the crustal torsional oscillations. Then, from the right panel of Fig.~\ref{fig4}, one can get the constraint on $L$ as $L\gtrsim 47.4$ MeV, if the central object of SGR 1806-20 is a neutron star with $M\le 1.8M_\odot$ and $R\le 14$ km \cite{SNIO12,SNIO13b}. Similarly, if the central object is a typical neutron star with $1.4M_\odot$ and 10 km, $L$ is constrained as $L\simeq 76.2$ MeV.

\begin{figure}[t]
\begin{center}
\includegraphics[width=6 cm]{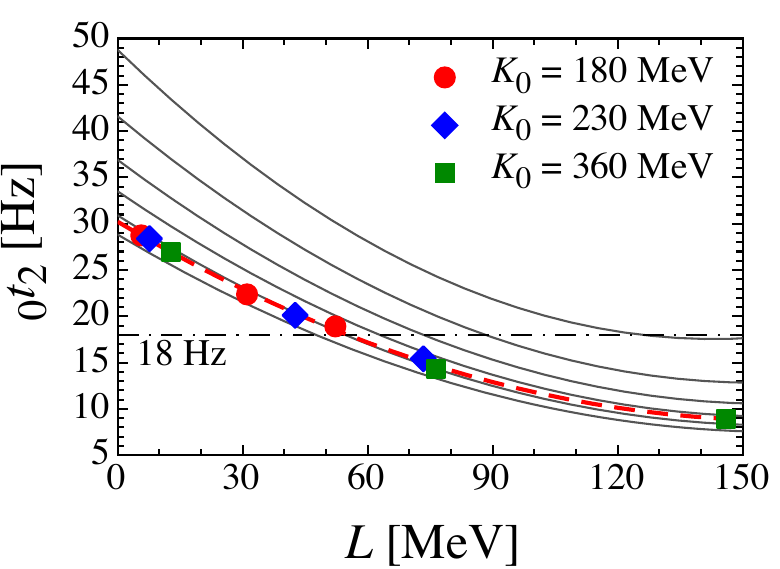}
\end{center}
\caption{ 
For a neutron star model with $1.8M_\odot$ and 14 km, the fundamental frequency of $\ell=2$ torsional oscillations is plotted as a function of $L$, as changing the ratio of $N_s/N_d$ with each 0.2 from 0 to 1, where $N_s/N_d$ is assumed to be constant inside the inner crust. Meanwhile, the marks denote the frequencies, using the density-dependent $N_s/N_d$ derived in \cite{Chamel12}, and the dashed line is their fitting with Eq. (\ref{eq:9}). Taken from \cite{SNIO13a}. \label{fig5}}
\end{figure}

Next, we take into account neutron superfluidity in the crustal torsional oscillations, i.e., $N_s/N_d\ne 0$. To see how the frequencies depend on $N_s/N_d$, we calculate ${}_0t_2$, assuming that $N_s/N_d$ is constant inside the inner crust. We confirm that ${}_0t_2$ for a given neutron star model can be well expressed as a function of $L$ using Eq.~(\ref{eq:9}) even for $N_s/N_d\ne 0$. In Fig.~\ref{fig5}, we show the $L$ dependence of ${}_0t_2$ for a neutron star model with $1.8M_\odot$ and 14 km as varying the ratio of $N_s/N_d$ with each 0.2 from 0 to 1 (the solid lines from bottom to top). The tendency of why the frequency increases with $N_s/N_d$ can also be understood with the dependence of frequency on the shear velocity. As mentioned above, the effective enthalpy, $\tilde{H}$, decreases with $N_s/N_d$, which leads to the increase of the shear velocity. So, as a result, one can expect that the frequency also increases since ${}_0t_\ell \sim v_s$. We note that the neutron star model discussed for the $L$ dependence of ${}_0t_2$ in Fig.~\ref{fig5} corresponds to the stellar model, which gives the lower boundary of the shaded region in Fig.~\ref{fig4}. That is, assuming that the central object in SGR 1806-20 is a neutron star with $M\le 1.8M_\odot$ and $R\le 14$ km, and also assuming that the lowest QPO frequency observed in SGR 1806-20 comes from ${}_0t_2$, one can get the constraint on $L$ as $L\gtrsim L_{min}$ with the lower limit of $L_{min}$, depending on $N_s/N_d$, where $L_{min}$ is determined as the intersection of the $L$ dependence of ${}_0t_2$ and 18 Hz (dot-dashed line) in Fig.~\ref{fig5}, i.e., $L_{min}$ increases with $N_s/N_d$ \cite{SNIO13a}.

In a realistic stellar model, $N_s/N_d$ should depend on the density. According to the band calculations in \cite{Chamel12}, the ratio of $N_s/N_d$ is only of the order of $10-30$ percent at $n_b\sim 0.01-0.4n_0$, even though the ratio of $N_s/N_d$ is still under debate, e.g., \cite{Cha05,CCH05,WP17}. The fundamental frequencies determined with this density dependence of $N_s/N_d$ are also shown in Fig.~\ref{fig5} with marks (and the dashed line for their fitting), which are close to the results obtained with $N_s/N_d=0.2$. Anyway, since the frequencies of torsional oscillations strongly depend on the ratio of $N_s/N_d$, it is quite important to understand the ratio of $N_s/N_d$ in a realistic situation.

\begin{figure}[t]
\begin{center}
\includegraphics[width=12 cm]{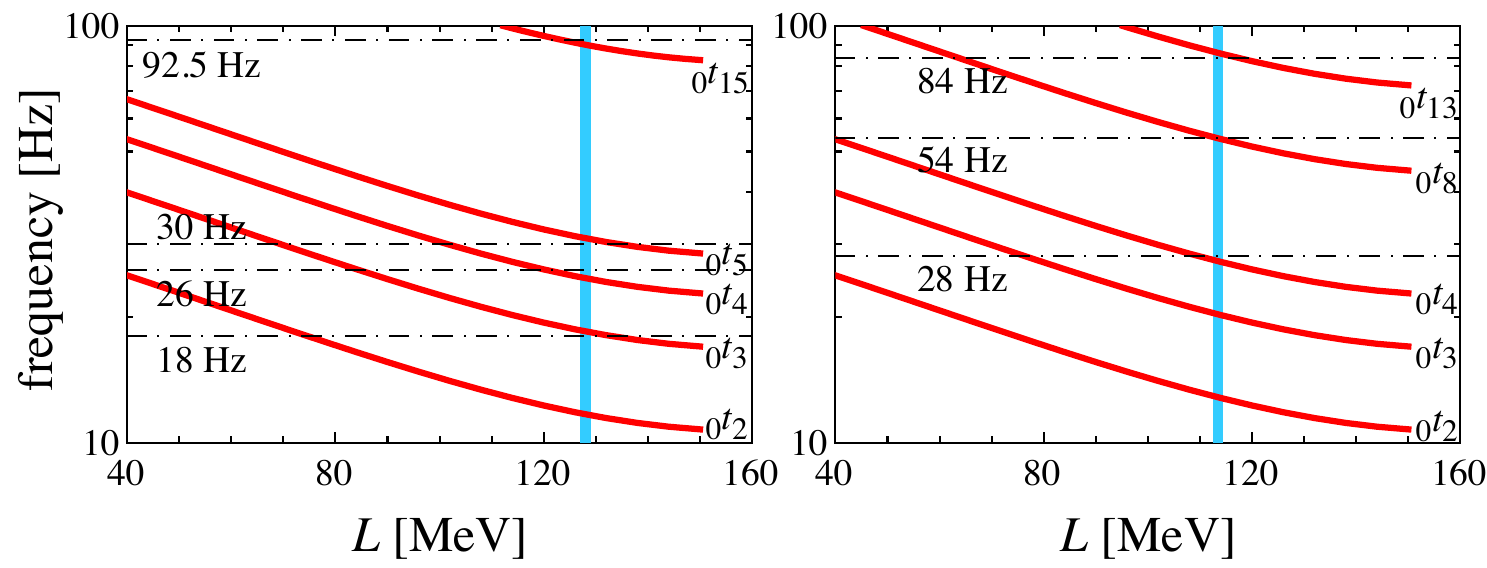}
\end{center}
\caption{ 
The fundamental frequencies of the torsional oscillations with several values of $\ell$ for a neutron star model with $1.4M_\odot$ and 12 km are shown as a function of $L$. The QPO frequencies observed in SGR 1806-20 and SGR 1900+14 are compared with them in the left and right panels.
Taken from \cite{SNIO13b}.\label{fig6}}
\end{figure}

In a similar way to the fundamental frequencies of the $\ell=2$ torsional oscillations, one can discuss the $L$ dependence of the $\ell$-th torsional oscillations. Actually, if one systematically examines the frequencies, one can find that ${}_0t_\ell$ is also well fitted as a function of $L$, which weakly depends on $K_0$, as
\begin{equation}
  {}_0t_\ell = c_\ell^{(0)} + c_\ell^{(1)} L_{100} + c_\ell^{(2)} L_{100}^2,
  \label{eq:10}
\end{equation}
where $c_\ell^{(i)}$ with $i=0,1,2$ are also the adjustable parameters depending on $M$ and $R$ (and $N_s/N_d$) \cite{SNIO13a}. Using this $L$ dependence of ${}_0t_\ell$, we will discuss the QPO frequencies observed in SGR 1806-20 and 1900+14, even though one should also consider the effect of the existence of pasta (see Sec. \ref{sec:5b} and \ref{sec:5c}, and also \cite{S11,PP16}). In Fig.~\ref{fig6}, we show the fundamental frequencies of torsional oscillations with specific values of $\ell$ for a neutron star model with $1.4M_\odot$ and 12 km, adopting the results in \cite{Chamel12} for $N_s/N_d$. From the left panel one can observe that the 18, 26, 30, and 92.5Hz QPOs in SGR 1806-20 are well identified with $\ell=3,4,5,15$ fundamental oscillations if $L\simeq 128.0$ MeV, while from the right panel one can observe that the 28, 54, and 84 Hz QPOs in SGR 1900+14 are with $\ell=4,8,13$ fundamental oscillations if $L\simeq 113.5$ MeV. We note that the other low-lying QPO frequencies discovered later can also be identified with the torsional oscillations with different values of $\ell$ in the same framework \cite{SIO16,SIO18,SIO19} (also see Sec. \ref{sec:5b} and \ref{sec:5c}).

\begin{figure}[t]
\begin{center}
\includegraphics[width=12 cm]{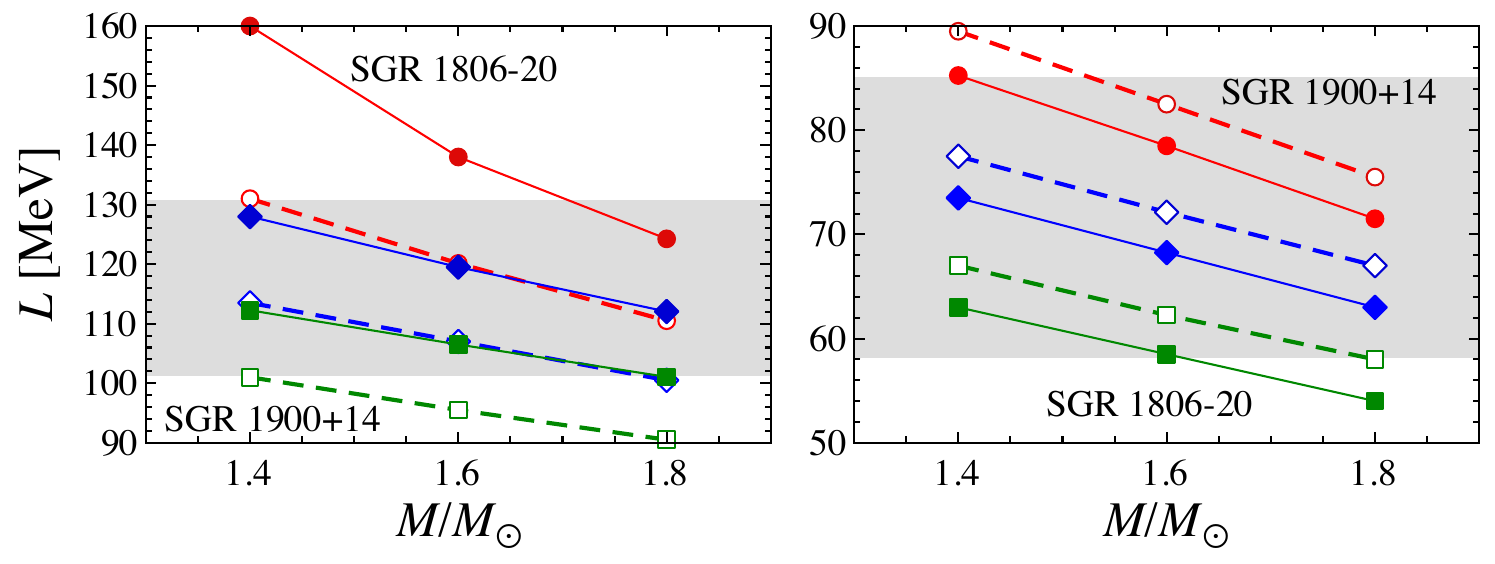}
\end{center}
\caption{ 
The values of $L$, with which the QPOs observed in SGR 1806-20 (SGR 1900+14) are well identified by the fundamental torsional oscillations, are shown with filled marks with solid lines (open marks with dashed lines) for different neutron star models, i.e., the circles, diamonds, and squares correspond to the stellar models with 10, 12, and 14 km radius. The left and right panels correspond to the identification shown in Fig.~\ref{fig6} and \ref{fig8}, respectively. The shaded region is the constraint on $L$, with which the QPOs observed in both SGR 1806-20 and 1900+14 are simultaneously explained by the torsional oscillations. 
Taken from \cite{SNIO13b}.\label{fig7}}
\end{figure}

Since the neutron star mass and radius in SGR 1806-20 and SGR 1900+14 are not constrained, it may be better to consider the stellar models with a typical range of mass and radius. In practice, if the stellar mass and radius are changed in a certain range, the resultant $L$ dependence of ${}_0t_\ell$ are shifted up and down in Fig.~\ref{fig6} with the same combination of the $\ell$-th oscillations. As a result, the optimal value of $L$, with which the QPOs observed in SGRs are well identified with the torsional oscillations, is also shifted left and right. In the left panel of Fig.~\ref{fig7}, the optimal values of $L$, with which the QPO frequencies observed in SGR 1806-20 (with filled marks) and in SGR 1900+14 (with open marks) are well identified as shown in Fig.~\ref{fig6}, are shown for various neutron star models. From this figure, one can see that $L$ should be in the range of $101.1\lesssim L\lesssim 160.0$ MeV for explaining SGR 1806-20 and $90.5\lesssim L\lesssim 131.0$ MeV for explaining SGR 1900+14, if the central objects of both SGRs are a neutron star with $1.4\le M/M_\odot \le 1.8$ and 10 km $\le R \le$ 14 km. Meanwhile, since the value of $L$ should be universal, i.e., independent of the astronomical events, one has to simultaneously explain both events, SGR 1806-20 and SGR 1900+14, with the same value of $L$. Thus, we get a more stringent constraint on $L$, i.e., $101.1\lesssim L \lesssim 131.0$ MeV \cite{SNIO13b}, which are shown in the left panel of Fig.~\ref{fig7} with the shaded region. Furthermore, since the symmetry energy at $n_b=n_0$, $S_0$, is approximately estimated as a function of $L$ \cite{OI03} through
\begin{equation}
  S_0 \approx 28\ {\rm MeV} + 0.075 L, \label{eq:S0}
\end{equation}
one can constrain $S_0$ to explain the QPOs observed in the SGRs as $35.6\lesssim S_0 \lesssim 37.8$ MeV.

\begin{figure}[t]
\begin{center}
\includegraphics[width=12 cm]{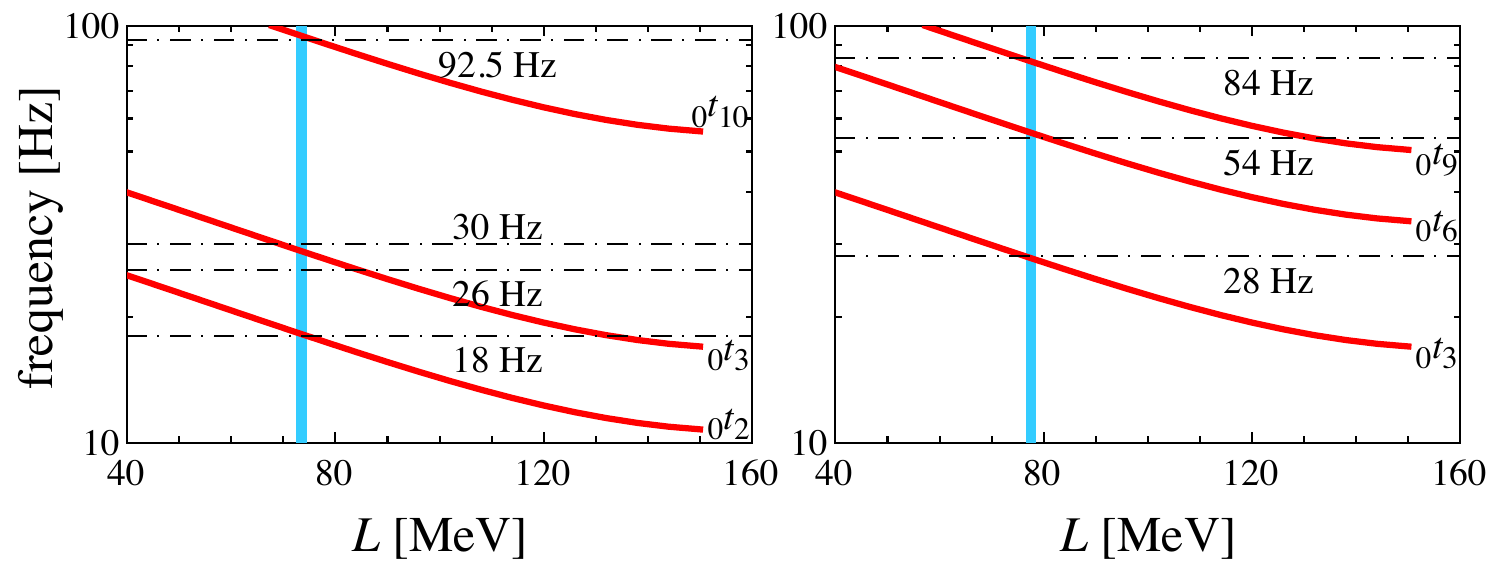}
\end{center}
\caption{ 
Alternative possible correspondence of the QPO frequencies observed in SGR 1806-20 (left panel) and SGR 1900+14 (right panel) with the fundamental torsional oscillations for a neutron star model with $1.4M_\odot$ and 12 km.
Taken from \cite{SNIO13b}.\label{fig8}}
\end{figure}

On the other hand, considering the constraints on $L$ obtained from terrestrial experiments whose fiducial value is $L\simeq 60\pm 20$ MeV \cite{Tsang12,New14,Li19}, the resultant constraint on $L$ may be too large, although a larger value of $L$ has also been reported \cite{PREXII,SPiRIT}. So, if any, it may be better to consider an alternative possible correspondence to explain the QPO frequencies observed in the SGRs with the fundamental torsional oscillations. As an alternative possible correspondence, we find that the 18, 30, and 92.5 Hz QPOs in SGR 1806-20 are identified with the $\ell=2,3,10$ fundamental torsional oscillations, while the 28, 54, and 84 Hz in SGR 1900+14 are with the $\ell = 3,6,9$ oscillations, as shown in Fig.~\ref{fig8} for a neutron star model with $1.4M_\odot$ and 12 km, where the value of $L$ should be $L\simeq 73.5$ MeV in SGR 1806-20 and $L\simeq 77.5$ MeV in 1900+14. We note that the 26 Hz QPO in SGR 1806-20 cannot be identified with this correspondence (see Sec. \ref{sec:5c} for this missing identification). Considering a typical neutron star model with $1.4\le M/M_\odot \le 1.8$ and 10 km $\le R \le$ 14 km, one can get the optimal values of $L$ as shown in the right panel of Fig.~\ref{fig7}. To simultaneously explain the QPOs observed in the SGRs, we obtain the constraint on $L$ as $58.0\lesssim L \lesssim 85.3$ MeV (shaded region in the right panel), assuming that the central object in the SGR 1806-20 and SGR 1900+14 is a neutron star with $1.4\le M/M_\odot \le 1.8$ and 10 km $\le R \le$ 14 km \cite{SNIO13b}. This constraint on $L$ additionally gives us the constraint on $S_0$ as $32.4\lesssim S_0 \lesssim 34.4$ MeV, using Eq. (\ref{eq:S0}). We note that the fundamental frequencies of $\ell$-th torsional oscillations excited inside the phase of spherical nuclei can be expressed as a function of $\ell$, $L$, stellar mass, and radius \cite{S16}.

\subsection{Torsional oscillations excited in the spherical and cylindrical nuclei}
\label{sec:5b}

Since in this review we simply consider that the shear modulus inside the phase of slablike nuclei is zero, as mentioned in Sec. \ref{sec:4}, the torsional oscillations in the region composed of spherical and cylindrical nuclei can be excited independently from those in the region composed of cylindrical-hole and spherical-hole nuclei. That is, torsional oscillations are excited independently in two layers across the phase of slablike (lasagna) nuclei as "a lasagna sandwich" \cite{SIO19}. In this subsection, we discuss the torsional oscillations excited in the outer layer, i.e., the phase of spherical and cylindrical nuclei, while those excited in the inner layer, i.e., the phase of cylindrical-hole and spherical-hole nuclei, will be discussed in the next subsection (Sec. \ref{sec:5c}).

Unlike the phase of spherical nuclei, the understanding of the ratio of $N_s/N_d$ in the phase of cylindrical nuclei is poor. So, we simply consider the extreme case, i.e., $N_s/N_d=0$ and 1 in the phase of cylindrical nuclei. On the other hand, we adopt the result obtained in \cite{Chamel12} for $N_s/N_d$ in the phase of spherical nuclei, as in the previous subsection. Then, one can determine the frequencies of torsional oscillations by solving an eigenvalue problem. 

\begin{figure}[t]
\begin{center}
\includegraphics[width=12 cm]{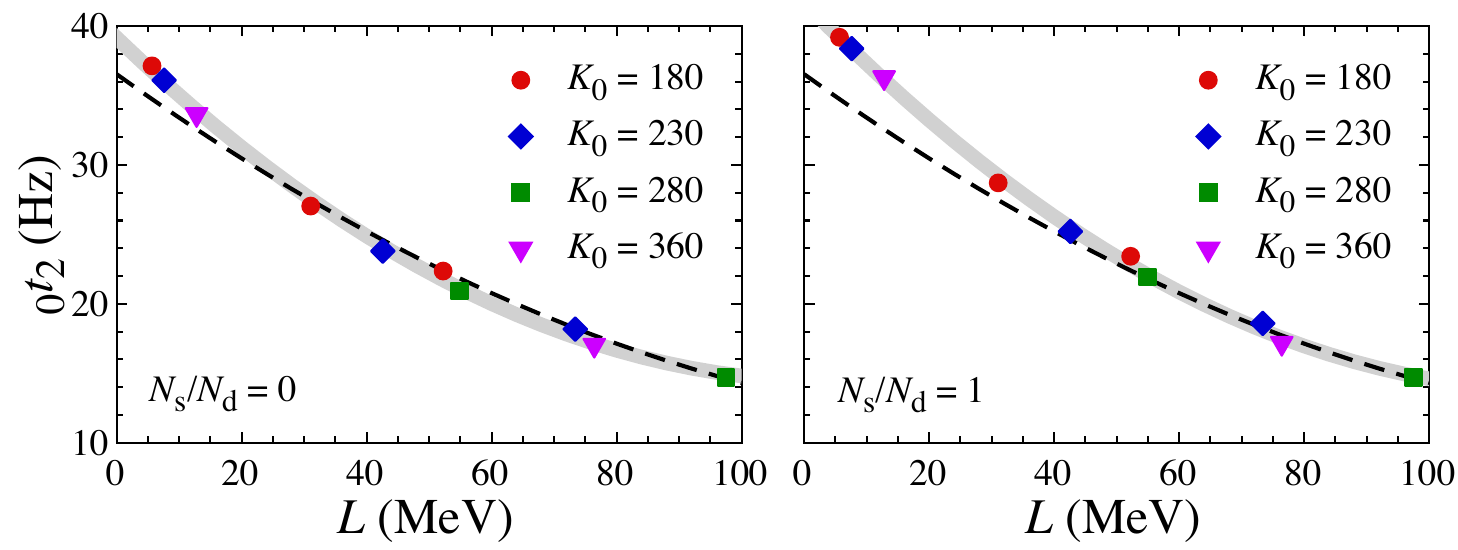}
\end{center}
\caption{ 
The fundamental frequencies of the $\ell=2$ torsional oscillations excited in the phase of spherical and cylindrical nuclei for a neutron star model with $1.4M_\odot$ and 12 km are shown with various marks. The left and right panels are results with $N_s/N_d=0$ and 1, respectively. The thick-solid line denotes the fitting with Eq. (\ref{eq:9}), while the dashed line denotes the $L$ dependence of ${}_0t_2$ exited in the phase of only spherical nuclei discussed in Sec. \ref{sec:5a}.
Taken from \cite{SIO18}.\label{fig9}}
\end{figure}

To see how the frequency of the torsional oscillation changes due to the existence of cylindrical nuclei, in Fig.~\ref{fig9}, we show ${}_0t_2$ as a function of $L$ for a neutron star model with $1.4M_\odot$ and 12 km, using the EOSs with some parameter sets listed in Table \ref{tab:SH-density} (see Ref. \cite{SIO18} for the concrete parameter sets adopted here), where the left and right panels correspond to the results with $N_s/N_d=0$ and 1, respectively, and the thick-solid line is the fitting with Eq. (\ref{eq:9}). For reference, we also show the $L$ dependence of ${}_0t_2$ excited in the phase of only spherical nuclei discussed in Sec.~\ref{sec:5a} with the dashed line. From this figure, one can observe that the modification in ${}_0t_2$ due to the existence of cylindrical nuclei appears only for a neutron star model with a lower value of $L$. This is because the phase of cylindrical nuclei as well as the other pasta phases becomes narrower with $L$ \cite{OI03,OI07,Oya23}, i.e., one can neglect the existence of cylindrical nuclei in the stellar model with larger $L$. In any case, the fundamental torsional oscillations hardly depend on $K_0$ and one can discuss the $L$ dependence through the fitting given by Eq.~(\ref{eq:10}). 

\begin{figure}[t]
\begin{center}
\includegraphics[width=12 cm]{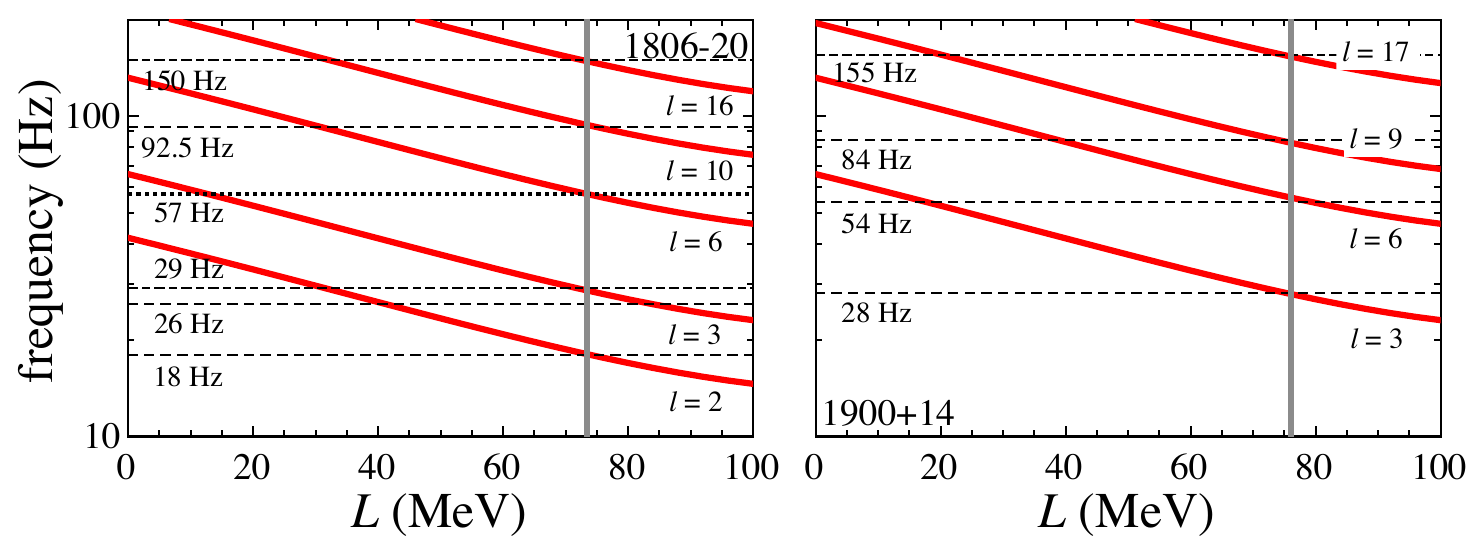}
\end{center}
\caption{ 
Correspondence of the low-lying QPO frequencies (except for the 26 Hz QPO) observed in SGR 1806-20 (left panel) and 1900+14 (right panel) with the several fundamental torsional oscillations for a neutron star model with $1.4M_\odot$ and 12 km, assuming that $N_s/N_d=1$ in the phase of cylindrical phase.
Taken from \cite{SIO18}.\label{fig10}}
\end{figure}

Using the $L$ dependence of ${}_0t_\ell$ with $N_s/N_d=1$ in the phase of cylindrical nuclei, in Fig.~\ref{fig10}, we compare the fundamental torsional oscillations with various $\ell$ for a neutron star model with $1.4M_\odot$ and 12 km to the low-lying QPO frequencies observed in SGR 1806-20 (left panel) and SGR 1900+14 (right panel), where the horizontal dashed and dotted lines denote the observed QPO frequencies and the solid lines denote the $L$ dependence of ${}_0t_\ell$. We note that we focus on only the correspondence of the observed low-lying QPOs except for the 26 Hz QPO in SGR 1806-20 here, because the optimal value of $L$ becomes larger than 100 MeV to identify all the observed low-frequency QPOs in terms of the crustal torsional oscillations, as discussed in Sec.~\ref{sec:5a}, which may be inconsistent with the constraint from existing nuclear experiments. We also note that the 57 Hz QPO additionally discovered in \cite{HHWG14} is taken into account this time. From this figure, one sees that the observed QPOs (except for the 26 Hz QPO) in SGR 1806-20 can be identified if $L\simeq 73.4$ MeV, and those in SGR 1900+14 can be identified if $L\simeq 76.1$ MeV. 

\begin{figure}[t]
\begin{center}
\includegraphics[width=12 cm]{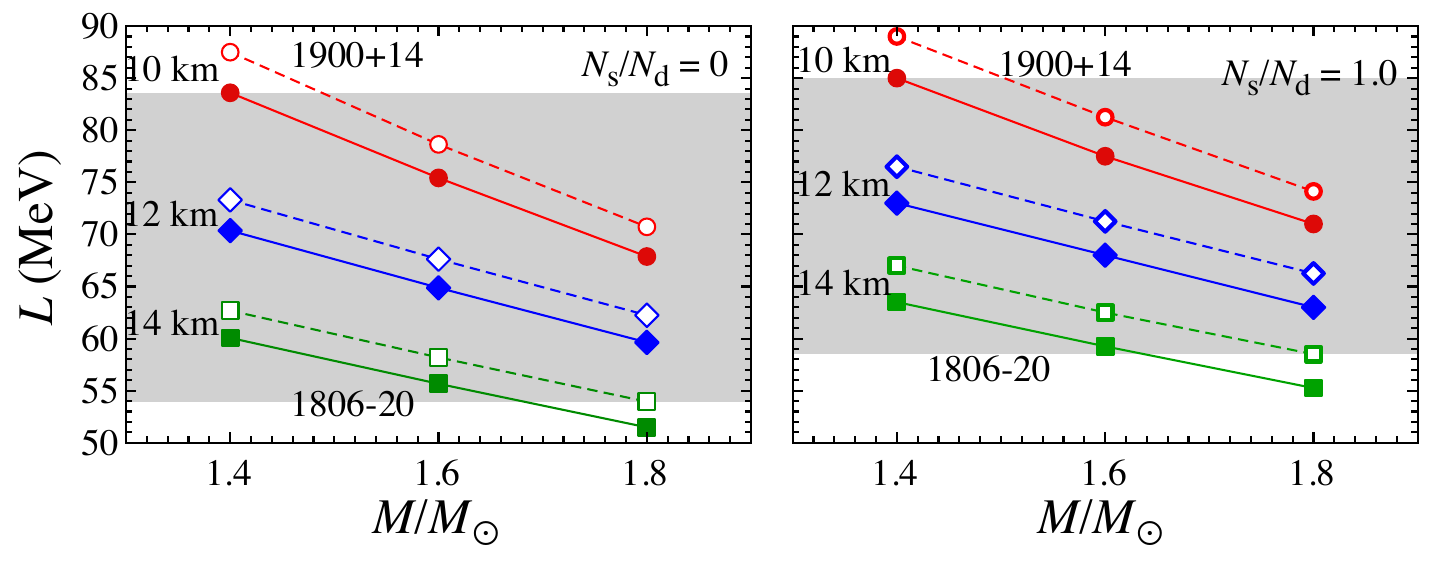}
\end{center}
\caption{ 
The optimal value of $L$ for identifying the low-lying QPO frequencies (except for the 26 Hz QPO) observed in SGR 1806-20 (filled marks with solid lines) and SGR 1900+14 (open marks with dashed lines) for various neutron star models, where the left and right panels are the results with $N_s/N_d=0$ and 1, respectively. 
Taken from \cite{SIO18}.\label{fig11}}
\end{figure}

Similarly, one can determine the optimal value of $L$ to identify the observed QPOs with the same set of the fundamental torsional oscillations shown in Fig~\ref{fig10} for various neutron star models with $N_s/N_d=0$ and 1 in the phase of cylindrical nuclei. Assuming that $1.4\le M/M_\odot \le 1.8$ and 10 km $\le R \le$ 14 km as a typical neutron star model, the optimal value of $L$ for identifying the QPO frequencies observed in the SGRs with the same correspondence as shown in Fig.~\ref{fig10}, are plotted in Fig.~\ref{fig11} for $N_s/N_d=0$ in the left panel and $N_s/N_d=1$ in the right panel. In this figure, the filled marks with solid lines correspond to the resultant values of $L$ for SGR 1806-20, while the open marks with dashed lines are for SGR 1900+14. Again, since the value of $L$ should be independent of the astronomical events, one has to simultaneously explain both events, SGR 1806-20 and SGR 1900+14, with a specific value of $L$. Thus, we can get the constraint on $L$ as $53.9 \lesssim L\lesssim 83.6$ MeV for $N_s/N_d=0$ and $58.1 \lesssim L \lesssim 85.1$ MeV for $N_s/N_d=1$, assuming that the central object of SGR 1806-20 and SGR 1900+14 is a neutron star with $1.4\le M/M_\odot \le 1.8$ and 10 km $\le R \le$ 14 km. These constraints on $L$ are shown with the shaded region in Fig.~\ref{fig11}. Namely, the uncertainty in $N_s/N_d$ in the phase of cylindrical nuclei makes only a little difference in the constraint on $L$, where the allowed $L$ lies in the range of $L = 53.9-85.1$ MeV, even if the uncertainty in $N_s/N_d$ in the phase of cylindrical nuclei is taken into account. This constraint on $L$ gives us the constraint of $S_0$ as $S_0\simeq 32.0-34.4$ MeV, using the correlation between $L$ and $S_0$ given by Eq.~(\ref{eq:S0}).

On the other hand, since the torsional oscillations are confined inside the phase of spherical and cylindrical nuclei, we can also discuss the overtone(s) of torsional oscillations. The $n$-th overtone frequencies of the $\ell$-th torsional oscillations, ${}_nt_\ell$, is theoretically estimated with the crust thickness (or the thickness of elastic region), $\Delta R$, as ${}_nt_\ell\sim v_s/\Delta R$~\cite{HC80,SA07}. Meanwhile, $\Delta R$ depends on the EOS parameters $L$ and $K_0$ mainly through the neutron chemical potential at the crust-core boundary \cite{SIO17b}, when the neutron star mass and radius are fixed. Thus, via the identification of the relatively high-frequency QPO observed in SGR 1806-20, i.e., 626.5 Hz, as the 1st overtone of crustal torsional oscillations, one may obtain information about the EOS parameters \cite{SNIO12,SIO18}.

\begin{figure}[t]
\begin{center}
\includegraphics[width=12 cm]{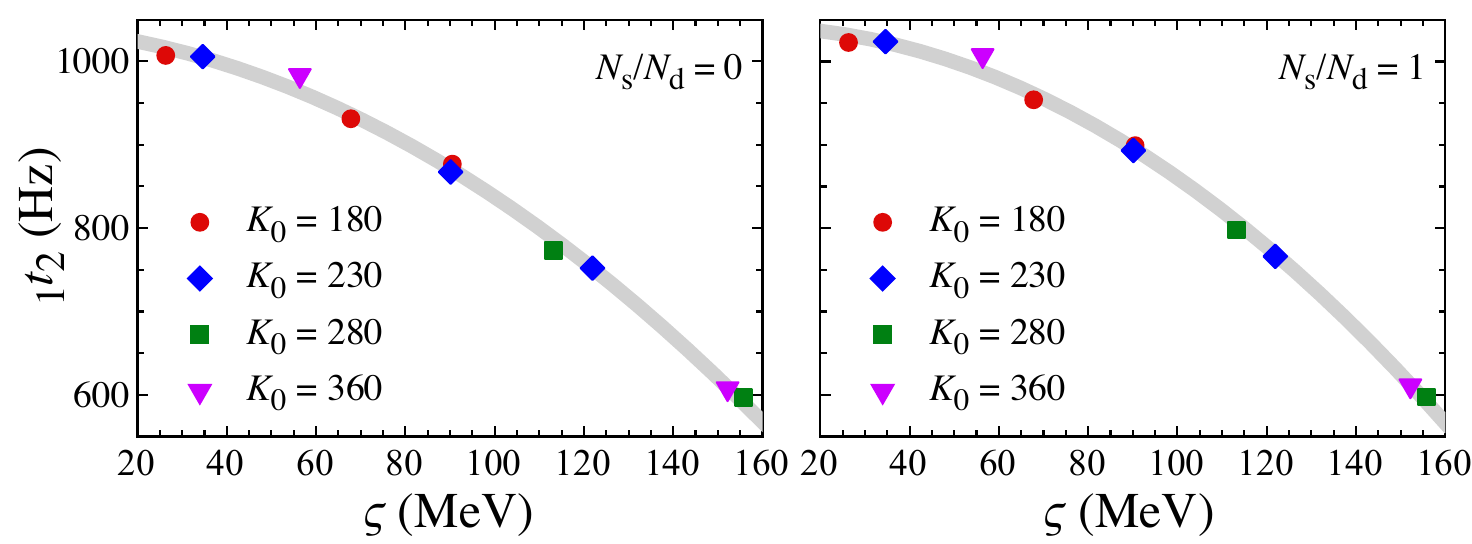}
\end{center}
\caption{ 
The 1st overtones of $\ell=2$ torsional oscillations for a neutron star model with $1.4M_\odot$ and 12 km constructed with various EOS parameters are shown as a function of $\varsigma$ defined by Eq.~(\ref{eq:12}), where the left and right panels respectively denote the results with $N_s/N_d=0$ and 1 in the phase of cylindrical nuclei. The thick-solid line denotes the fitting formula given by Eq.~(\ref{eq:13}).
Taken from \cite{SIO18}.\label{fig12}}
\end{figure}

Since ${}_nt_\ell$ depends on $K_0$ and $L$ through $\Delta R$, it is of great use to find a parameter constructed by a combination of $K_0$ and $L$, which can characterize ${}_nt_\ell$. To this end, assuming the combination of $(K_0^iL^j)^{1/(i+j)}$ with integer numbers $i$ and $j$, we finally find an appropriate combination of $K_0$ and $L$ as
\begin{equation}
  \varsigma \equiv (K_0^4L^5)^{1/9}. 
  \label{eq:12}
\end{equation}
In Fig.~\ref{fig12}, we show the 1st overtones of $\ell=2$ torsional oscillations for a neutron star model with $1.4M_\odot$ and 12 km constructed with various EOS parameters, where the thick-solid line denotes the fitting formula for the 1st overtones of the torsional oscillations given by \cite{SIO18,SKS23}
\begin{equation}
  {}_nt_\ell = d_{\ell n}^{(0)} + d_{\ell n}^{(1)}\varsigma + d_{\ell n}^{(2)}\varsigma^2, 
  \label{eq:13}
\end{equation}
where $d_{\ell n}^{(i)}$ for $i=0,1,2$ are the adjustable parameters depending on $M$, $R$, and $N_s/N_d$ in the phase of cylindrical nuclei. We will, hereafter, discuss the $\varsigma$ dependence of ${}_nt_\ell$. 

It is well known that the overtone frequencies of torsional oscillations weakly depend on $\ell$, unlike the fundamental oscillations \cite{HC80}. As an example, in the left panel, we show the 1st overtones of the $\ell=2$ (solid lines) and 10 (dashed lines) torsional oscillations as a function of $\varsigma$ for a neutron star models with $(M,R)=(1.4M_\odot, 10\ {\rm km})$, $(1.6M_\odot, 12\ {\rm km})$, and $(1.8M_\odot, 14\ {\rm km})$ with $N_s/N_d=1$ in the phase of cylindrical nuclei. From this figure, one confirms that ${}_1t_\ell$ weakly depends on $\ell$, while one also finds that ${}_1t_\ell$ strongly depends on the stellar models. So, hereafter, in this review, we focus only on ${}_nt_2$ to discuss the $\varsigma$ dependence of the overtones. In addition, in the right panel, we show the $\varsigma$ dependence of ${}_1t_2$ for the $1.4M_\odot$ neutron star models with different radii with $N_s/N_d=1$ in the phase of cylindrical nuclei together with the 626.5 Hz QPO observed in SGR 1806-20. From this figure, one can observe that the optimal values of $\varsigma$ are 178.5, 149.7, and 107.1 MeV for $1.4M_\odot$ neutron stars of $R = 10$, 12, and 14 km, respectively, if one identifies the 626.5 Hz QPO with the 1st overtone. 

\begin{figure}[t]
\begin{center}
\includegraphics[width=12 cm]{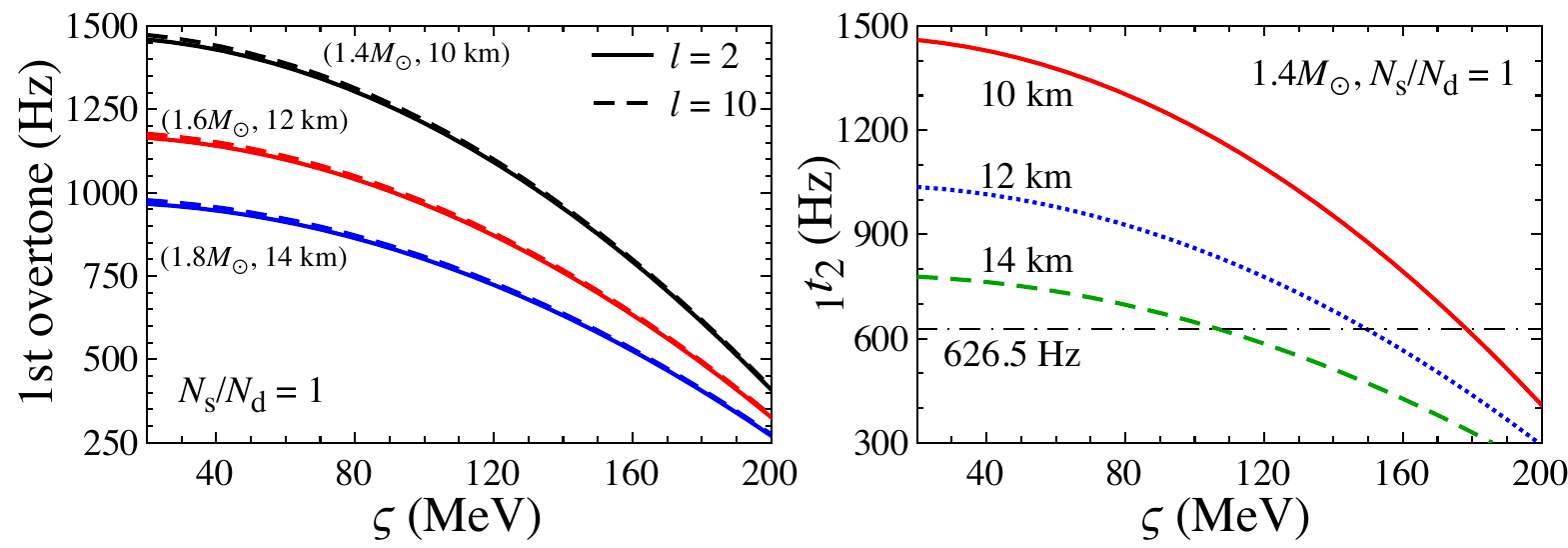}
\end{center}
\caption{ 
In the left panel, we show the 1st overtones of the $\ell=2$ (solid lines) and 10 (dashed lines) torsional oscillations as a function of $\varsigma$ for a neutron star models with $(M,R)=(1.4M_\odot, 10\ {\rm km})$, $(1.6M_\odot, 12\ {\rm km})$, and $(1.8M_\odot, 14\ {\rm km})$ with $N_s/N_d=1$ in the phase of cylindrical nuclei. In the right panel, the $\varsigma$ dependence of the 1st overtone of the $\ell=2$ torsional oscillations for $1.4M_\odot$ neutron star models with $R=10$ km (solid line), 12 km (dotted line), and 14 km (dashed line) with $N_s/N_d=1$ in the phase of cylindrical nuclei is compared to the 626.5 Hz QPO observed in SGR 1806-20 (dot-dashed line). 
Taken from \cite{SIO18}.\label{fig13}}
\end{figure}   

\begin{figure}[t]
\begin{center}
\includegraphics[width=12 cm]{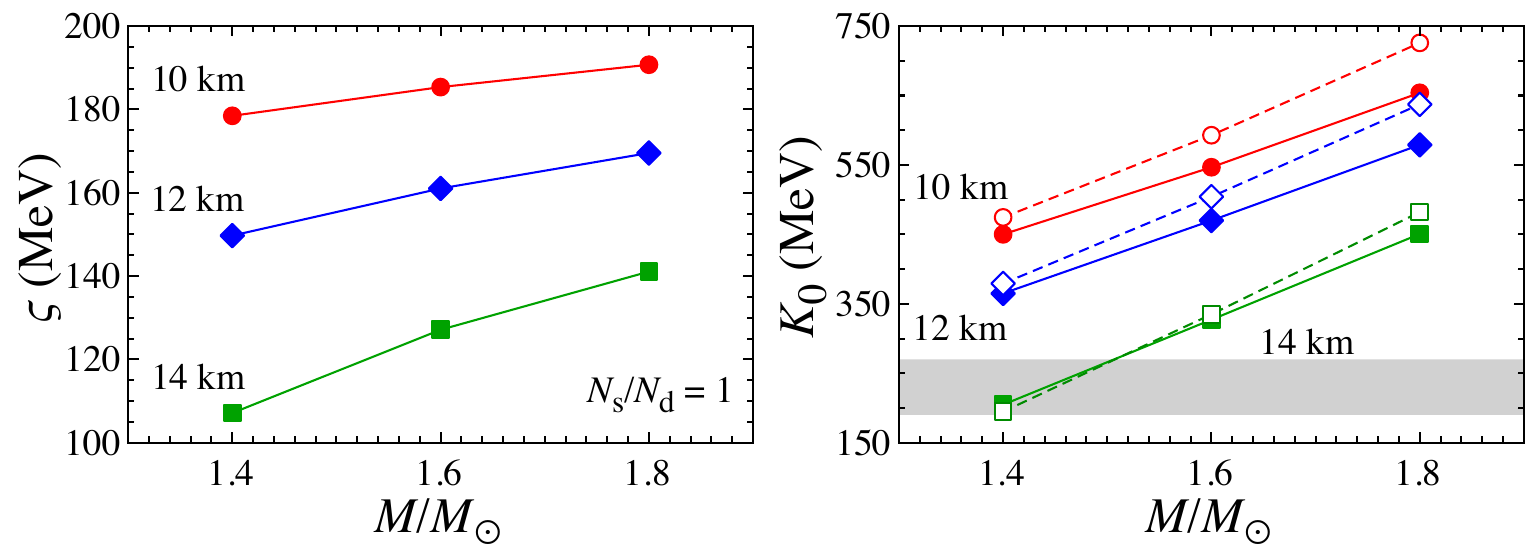}
\end{center}
\caption{ 
In the left panel, the optimal value of $\varsigma$ to identify the 626.5 Hz QPO observed in SGR 1806-20 with the 1st overtone of the $\ell=2$ torsional oscillations are plotted for various neutron star models with $1.4\le M/M_\odot \le 1.8$ and 10 km $\le R \le$ 14 km for $N_s/N_d=1$ in the phase of cylindrical nuclei. In the right panel, the constraint on $K_0$ obtained by combining the optimal values of $\varsigma$ shown in the left panel (and also those obtained with $N_s/N_d=0$) and the optimal values of $L$ shown in Fig.~\ref{fig11} is plotted for each stellar model, where filled marks with solid lines (open marks with dashed lines) denote the results for $N_s/N_d=1$ (0) in the phase of cylindrical nuclei. Meanwhile, the shaded region denotes the experimental constraint on $K_0$ obtained in \cite{KM13}, i.e., $K_0=230\pm 40$ MeV.
Taken from~\cite{SIO18}.\label{fig14}}
\end{figure}

In a similar way, one can obtain the optimal values of $\varsigma$ for various neutron star models. In the left panel of Fig.~\ref{fig14}, the resultant optimal values of $\varsigma$ are plotted for neutron star models with $1.4\le M/M_\odot \le 1.8$ and 10 km $\le R \le$ 14 km for $N_s/N_d=1$ in the phase of cylindrical nuclei. One can observe that for each $R$, the optimal $\varsigma$ increases with $M$. This behavior comes from the fact that $\Delta R/R$ (or $\Delta R$ with fixed $R$) decreases with the compactness, $M/R$ \cite{SIO17b}, which leads to the increases of ${}_1t_2$ (through ${}_1t_2\sim v_s/\Delta R$) and that a optimal value of $\varsigma$ becomes larger.

Furthermore, we also derive constraint on $K_0$ through the definision of $\varsigma$ given by Eq.~(\ref{eq:12}), i.e., $K_0=(\varsigma^9/L^5)^{1/4}$, using the optimal values of $\varsigma$ for identifying the 626.5 Hz QPO with the 1st overtone frequency shown in the left panel of Fig.~\ref{fig14} and the optimal values of $L$ for identifying the low-lying QPO frequencies shown in Fig.~\ref{fig11}. In the right panel of Fig.~\ref{fig14}, such constraints on $K_0$ are plotted for various neutron star models, where the filled marks with solid lines denote the results for $N_s/N_d = 1$, while the open marks with dashed lines denote the results for $N_s/N_d = 0$. In the same panel, we also show the experimental constraint on $K_0$, i.e., $K_0=230\pm 40$ MeV \cite{KM13}. Adopting this constraint on $K_0$ as a typical one, although it may still be model dependent, e.g., \cite{SSM14}, from the right panel of Fig.~\ref{fig14} one can observe that the neutron star models with $M\simeq 1.4-1.5 M_\odot$ for $R=14$ km and presumably $M\simeq 1.2-1.4 M_\odot$ for $R=13$ km are favored by the QPOs observed in SGR 1806-20 up to 626.5 Hz (except for the 26 Hz QPO). That is, a central object in SGR 1806-20 would have a relatively low mass and large radius. We note in passing that a neutron star model with still lower mass and smaller radius than that mentioned above might be acceptable from the right panel of Fig.~\ref{fig14}, but one has to assume a larger value of $L$ to construct such a stellar model, which would be presumably inconsistent with the systematic analysis of the mass-radius relation for low-mass neutron stars \cite{SIOO14}.

Finally, adopting the resultant constraint on the neutron star model of SGR 1806-20, i.e., $M\simeq 1.4-1.5 M_\odot$ for $R=14$ km and $M\simeq 1.2-1.4 M_\odot$ for $R=13$ km, the constraints on $L$ shown in Fig.~\ref{fig11} are dramatically improved. Namely, the optimal value of $L$ should be $L \simeq 62-73$ MeV for $N_s/N_d=1$ and $L \simeq 58-70$ MeV for $N_s/N_d=0$ in the phase of cylindrical nuclei. Therefore, we obtain the constraint on $L$ as $L\simeq 58-73$ MeV independently of the uncertainty in $N_s/N_d$ in the phase of cylindrical nuclei, which is consistent with the existing constraint on $L$, e.g., \cite{Tsang12,New14,Li19}. Using the correlation between $L$ and $S_0$ (Eq.~(\ref{eq:S0})), one can estimate the corresponding value of $S_0$ as $S_0\simeq 32.4-33.5$ MeV.

\subsection{Torsional oscillations excited in the cylindrical-hole and spherical-hole nuclei}
\label{sec:5c}

Through Sec.~\ref{sec:5a} and \ref{sec:5b}, we discussed that the QPO frequencies observed in SGR 1806-20 and SGR 1900+14 are well identified with the torsional oscillations excited in the phase of spherical and cylindrical nuclei. In particular, by identifying the 626.5 Hz QPO with the 1st overtone of the torsional oscillations, we showed the possibility that a central object in SGR 1806-20 would be a neutron star with a relatively low mass and large radius, and derived the constraint on $L$ more severe. However, still, we have a missing piece in the identification of the observed QPO frequencies, i.e.,  the 26 Hz QPO observed in SGR 1806-20. In this subsection, we discuss the possibility of identifying the 26 Hz QPO (and the QPOs additionally discovered in \cite{MCS19}) in SGR 1806-20 with the torsional oscillations excited in the phase of cylindrical-hole and spherical-hole nuclei, keeping the consistency with the identification discussed in Sec.~\ref{sec:5b}.

\begin{figure}[t]
\begin{center}
\includegraphics[width=12 cm]{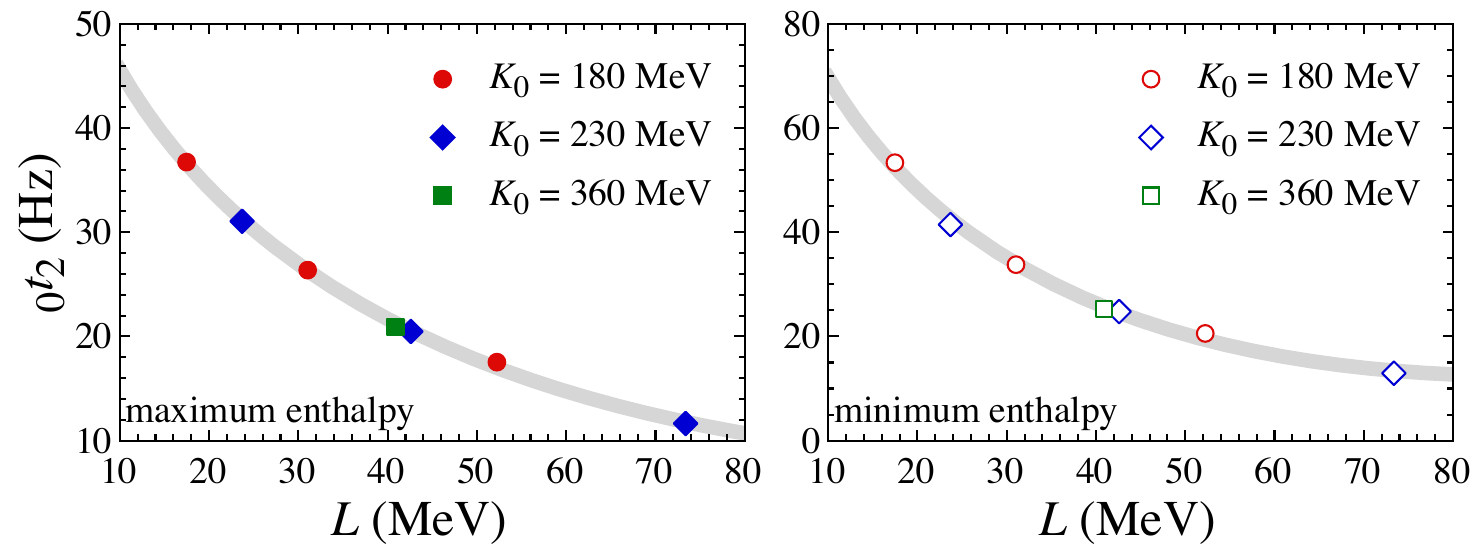}
\end{center}
\caption{ 
The fundamental frequencies of the $\ell=2$ torsional oscillations excited in the phase of cylindrical-hole and spherical-hole nuclei for a neutron star model with $1.4M_\odot$ and 12 km. The thick-solid line corresponds to the fitting formula given by Eq.~(\ref{eq:14}). The left and right panels respectively correspond to the results with ${\cal R}=1$ (maximum enthalpy) and ${\cal R}=0$ (minimum enthalpy).
Taken from \cite{SIO19}.\label{fig15}}
\end{figure}   

To determine the frequencies of torsional oscillations excited in the phase of cylindrical-hole and spherical-hole nuclei, one has to know the value of ${\cal R}$ in Eq.~(\ref{eq:8}) in the phases of cylindrical-hole and spherical-hole nuclei, but it is still quite uncertain. So, here we simply consider the extreme cases, i.e., ${\cal R}=1$ for maximum enthalpy and 0 for minimum enthalpy. In Fig.~\ref{fig15}, we show the fundamental frequencies of the $\ell=2$ torsional oscillations excited in the phase of cylindrical-hole and spherical-hole nuclei for a neutron star model with $1.4M_\odot$ and 12 km, using the EOSs with some parameter sets listed in Table~\ref{tab:SH-density} (see Ref. \cite{SIO19} for the concrete parameter sets adopted here), where the left and right panels correspond to the results with ${\cal R}=1$ and 0 in the phase of cylindrical-hole and spherical-hole nuclei, respectively. From this future, we find that the fundamental frequencies excited in the phase of cylindrical-hole and spherical-hole nuclei weakly depend on $K_0$ and the dependence on $L$ is well fitted with the functional form given by
\begin{equation}
  {}_0t_\ell = \tilde{c}_{\ell}^{(0)} + \tilde{c}_{\ell}^{(1)}\sqrt{L} + \tilde{c}_{\ell}^{(2)}L, 
  \label{eq:14}
\end{equation}
where $\tilde{c}_{\ell}^{(i)}$ for $i=0,1,2$ are the adjustable parameters depending on $M$, $R$, and ${\cal R}$ in the phase of cylindrical-hole and spherical-hole nuclei \cite{SIO19}. Using this $L$ dependence of ${}_0t_\ell$ excited in the phase of cylindrical-hole and spherical-hole nuclei, we will see the correspondence of the observed QPOs.

\begin{figure}[t]
\begin{center}
\includegraphics[width=13 cm]{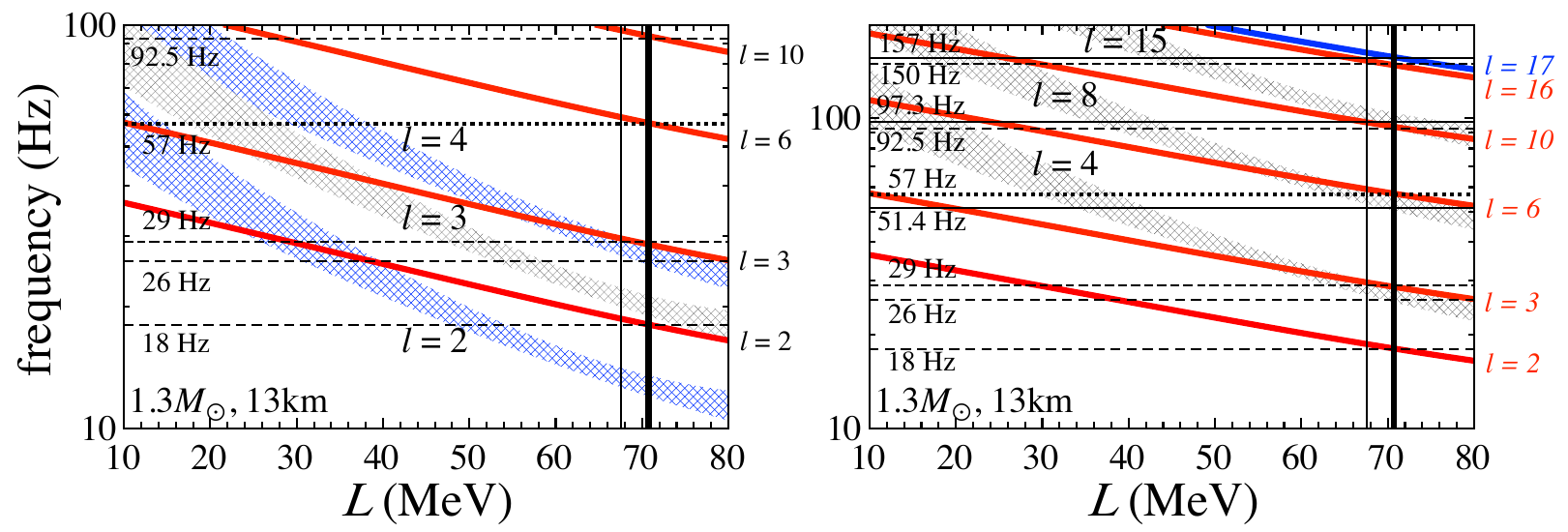}
\end{center}
\caption{ 
The correspondence of the QPO frequencies observed in SGR 1806-20 and the fundamental frequencies of torsional oscillations with various values of $\ell$ excited in the phase of spherical and cylindrical nuclei with $N_s/N_d=1$ in the cylindrical nuclei (solid lines) and the phase of cylindrical-hole and spherical-hole nuclei for $0\le {\cal R}\le 1$ (shaded regions) for a neutron star model with $1.3M_\odot$ and 13 km. We focus on only the low-lying QPOs in the left panel, while in the right panel we also consider the QPOs additionally discovered in \cite{MCS19} together with the QPOs considered in the left panel. The vertical lines denote the optimal values of $L$ to identify the low-lying QPO frequencies observed in SGR 1806-20 except for the 26 Hz QPO with the torsional oscillations excited in the phase of spherical and cylindrical nuclei discussed in Sec.~\ref{sec:5b}, where the thick line ($L \simeq 70.8$ MeV) and thin line ($L \simeq 67.5$ MeV) respectively correspond to the optimal values of $L$ with $N_s/N_d=1$ and 0 in the phase of cylindrical nuclei. 
Taken from \cite{SIO19}.\label{fig16}}
\end{figure}   

In the left panel of Fig.~\ref{fig16}, we show the identification of the low-lying QPOs observed in SGR 1806-20 with the fundamental frequencies of torsional oscillations excited in the phase of spherical and cylindrical nuclei and the phase of cylindrical-hole and spherical-hole nuclei for a neutron star model with $1.3M_\odot$ and 13 km. As shown in the left panel of Fig.~\ref{fig10}, the 18, 29, 57, 92.5 Hz QPOs can be identified with the fundamental frequencies of the $\ell=2$, 3, 6, 10 torsional oscillations excited in the phase of spherical and cylindrical nuclei. With this correspondence, we find the optimal values of $L$ as $L \simeq 70.8$ MeV for $N_s/N_d=1$ and $L \simeq 67.5$ MeV for $N_s/N_d=0$ in the phase of cylindrical nuclei. We note that in Fig.~\ref{fig16} the solid lines denote the fundamental frequencies of torsional oscillations excited in the phase of spherical and cylindrical nuclei with $N_s/N_d=1$ in the phase of cylindrical nuclei. On the other hand, we also show the fundamental frequencies of $\ell=2,3,4$ torsional oscillations excited in the phase of cylindrical-hole and spherical-hole nuclei by the shaded regions, assuming that $0\le {\cal R}\le 1$. From this figure, one can observe that the 26 Hz QPO, which cannot be identified with the torsional oscillations in the phase of spherical and cylindrical nuclei, can be identified with the $\ell=4$ fundamental torsional oscillations excited in the phase of cylindrical-hole and spherical-hole nuclei consistently with the optimal value of $L$ given by the identification of the other QPOs with the torsional oscillations in the phase of spherical and cylindrical nuclei.

In addition, using this double layer model (lasagna sandwich model), we find that one can identify the QPOs originally discovered in SGR 1806-20 together with the QPOs additionally discovered by a Bayesian procedure, e.g., 51.4, 97.3, and 157 Hz QPOs \cite{MCS19}, as shown in the right panel of Fig.~\ref{fig16}. That is, as shown in the left panel of Fig.~\ref{fig10}, the 18, 29, 57, 92.5, and 150 Hz QPOs can be identified with the fundamental frequencies of the $\ell = 2, 3, 6, 10$, and 16 torsional oscillations excited in the phase of spherical and cylindrical nuclei. In a similar way, the  157 Hz QPO can be identified with the $\ell=17$ fundamental torsional oscillations in the phase of spherical and cylindrical nuclei, while the 26, 51.4, and 97.3 Hz QPOs are the $\ell=4,8$, and 15 fundamental torsional oscillations in the phase of cylindrical-hole and spherical-hole nuclei.

\subsection{Constraint on a neutron star model for GRB 200415A}
\label{sec:5d}

So far, we have considered the correspondence between the crustal torsional oscillations and the QPO frequencies observed in giant flares. In addition to these observations, another magnetar flare, GRB 200415A, was also detected in the direction of the NGC253 galaxy, where several high-frequency QPOs with varying significance were found at 836, 1444, 2132, and 4250 Hz \cite{C21}. We note that only high-frequency QPOs have been detected from this event due to the short duration of the observation interval. So, considering the dynamical time of neutron stars, it may be possible to identify these observed QPOs with neutron star oscillations other than the torsional oscillations, such as the fundamental ($f$-), gravity ($g_i$-), or shear ($s_i$-) modes (e.g., see Fig.~\ref{fig:QPO-is}). Nevertheless, since these QPOs come from the magnetar flare, here we consider the identification with the crustal torsional oscillations, which is the same framework as discussed in the case of the QPOs observed in giant flares (Sec.~\ref{sec:5a},~\ref{sec:5b}, and~\ref{sec:5c}). 

\begin{figure}[t]
\begin{center}
\includegraphics[width=12 cm]{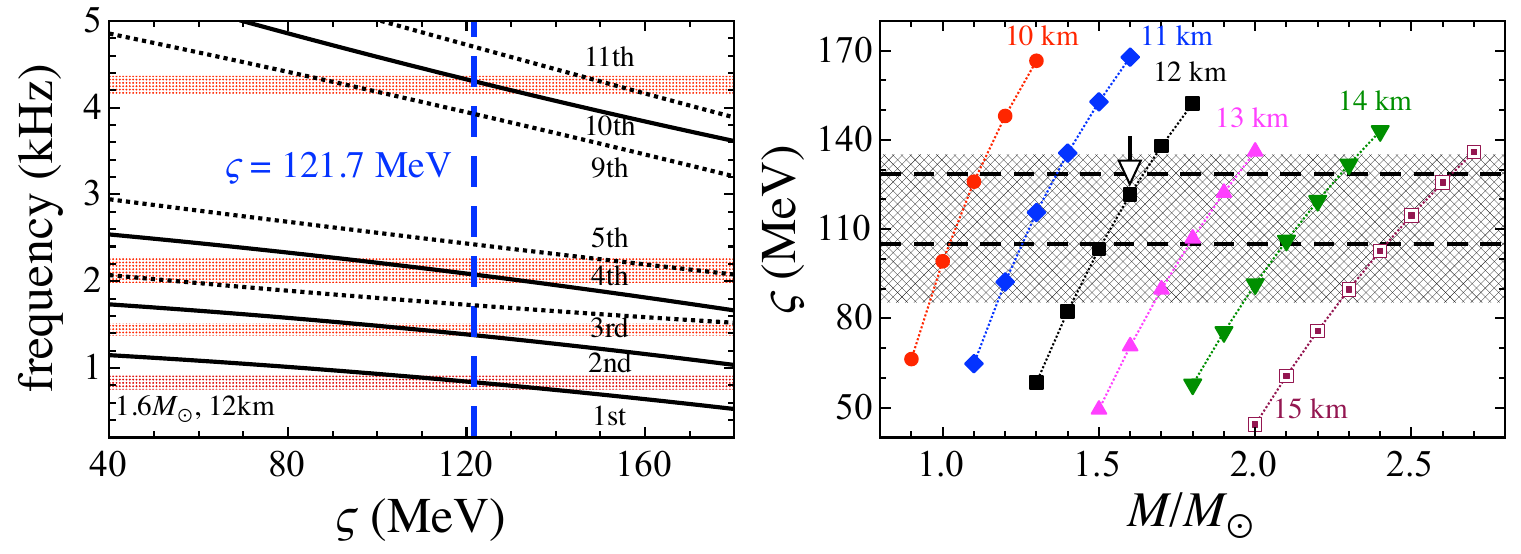}
\end{center}
\caption{ 
In the left panel, we show the identification of four QPO frequencies observed in GRB 200415A \cite{C21} (shaded horizontal regions) with several overtones of torsional oscillations excited in the phase of spherical and cylindrical nuclei for a neutron star model with $1.6M_\odot$ and 12 km with $N_s/N_d=0$ in the phase of cylindrical nuclei, where $\varsigma=121.7$ is the optimal value with this correspondence. In the right panel, we show the optimal values of $\varsigma$ to identify the four QPOs in GRB 200415A with the same combinations of overtones as in the left panel for various neutron star models. The shaded region is the range of $\varsigma=85.3-135.1$ MeV estimated with the fiducial value of $L=60\pm 20$ MeV and $K_0=240\pm 20$ MeV, while the region in the dashed lines is the range of $\varsigma_{QPO}=104.9-128.4$ MeV with the optimal values of $L$ discussed in Sec.~\ref{sec:5b}, i.e., $L=58-73$ MeV, together with $K_0=240\pm 20$ MeV. The arrow in the right panel indicates the neutron star model considered in the left panel.
Taken from \cite{SKS23}.\label{fig17}}
\end{figure}

Since the observed QPOs are too high to identify the fundamental torsional oscillations, we consider identifying them with the overtones of torsional oscillations, as in Sec.~\ref{sec:5b}. Using the fitting of the overtone frequencies given by Eq.~(\ref{eq:13}), we plot the $\varsigma$ dependence of ${}_nt_2$ in the left panel of Fig.~\ref{fig17} for a neutron star model with $1.6M_\odot$ and 12 km with $N_s/N_d=0$ in the phase of cylindrical nuclei, where the horizontal shaded regions denote the QPOs observed in GRB 200415A. We note that the overtone frequencies weakly depend on the ratio of $N_s/N_d$ in the phase of cylindrical nuclei~\cite{SKS23}, in this subsection, we only consider the case with $N_s/N_d=0$ in the phase of cylindrical nuclei. From this figure, one can observe that the four QPOs can be identified well with the 1st, 2nd, 4th, and 10th overtones of torsional oscillations, which tells us the optimal value of $\varsigma$ is 121.7 MeV. 

\begin{figure}[t]
\begin{center}
\includegraphics[width=12 cm]{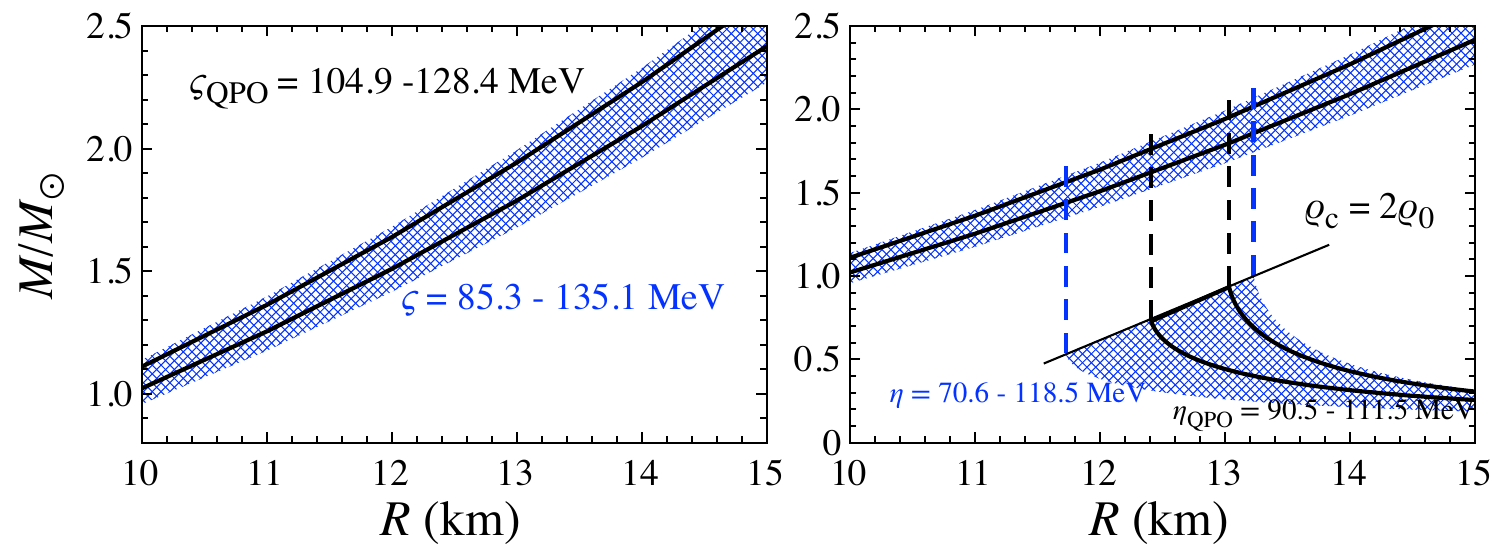}
\end{center}
\caption{ 
In the left panel, the constraint on the neutron star mass and radius for GRB 200415A obtained from the optimal values of $\varsigma$ to identify the high-frequency QPOs with 1st, 2nd, 4th, and 10th overtones of torsional oscillations, adopting the range of $\varsigma=85.3-135.1$ MeV estimated with the fiducial value of $L$ and $K_0$ (shaded region) or $\varsigma_{QPO}=104.9-128.4$ MeV with the optimal value of $L$ to identify the magnetar QPOs (bound by solid lines). In the right panel, we show the constraint on the neutron star mass and radius obtained from the QPOs in GRB 200415A (the same as the left panel) together with those estimated with the mass formula for a low-mass neutron star as a function of $\eta = (K_0L^2)^{1/3}$ derived in \cite{SIOO14} (the right-bottom region), assuming that the central density of neutron stars should be less than twice the nuclear saturation density. 
Taken from \cite{SKS23}.\label{fig18}}
\end{figure}

In a similar way, one can determine the optimal values of $\varsigma$ for various neutron star models. In practice, the optimal values of $\varsigma$ determined with the same combination of the overtones as shown in the left panel of Fig.~\ref{fig17} are plotted in the right panel of Fig.~\ref{fig17}. In particular, the neutron star model considered in the left panel is indicated with the arrow. Meanwhile, the value of $\varsigma$ is also estimated with the fiducial value of $L$ and $K_0$, i.e., $\varsigma=83.5-135.1$ MeV with $L=60\pm 20$ MeV and $K_0=240\pm 20$ MeV \cite{SKC06}, or with the optimal value of $L$ obtained in Sec.~\ref{sec:5b} to identify the QPO frequencies observed in the giant flares with the torsional oscillations, such as $L=58-73$ MeV, with the fiducial value of $K_0$, i.e., $\varsigma_{QPO}=104.9-128.4$ MeV. These estimations of $\varsigma$ are also shown in the right panel of Fig.~\ref{fig17} with the shaded region and the enclosed region with the dashed lines. That is, considering these estimations, for example, the stellar model with $1.5M_\odot$ and 11 km can be excluded. In consequence, we can get the constraint on the neutron star mass and radius as shown in the left panel of Fig.~\ref{fig18}, where the shaded region corresponds to the results with $\varsigma=83.5-135.1$ MeV, while the bound region by solid lines is those with $\varsigma_{QPO}=104.9-128.4$ MeV.

One may make the constraint on the mass and radius of the neutron star model for GRB 200415A more severe, if one knows the EOSs (or mass-radius relations) corresponding to the estimations of $\varsigma$. For this purpose, the mass formula for a low-mass neutron star \cite{SIOO14} must be crucial. The low-mass neutron star structures may strongly depend on the nuclear saturation parameters, because the central density for such an object is very low. In fact, the mass and gravitational redshift, $z$, defined by $z=1/\sqrt{1-2M/R}-1$ can be well expressed as a function of the stellar central density and a new parameter given by 
\begin{equation}
  \eta \equiv (K_0L^2)^{1/3}, 
  \label{eq:15}
\end{equation}
such that
\begin{gather}
  M/M_\odot = 0.371 - 0.820u_c + 0.279u_c^2 - (0.593 - 1.25u_c + 0.235u_c^2)\eta_{100}, \label{eq:16} \\
  z = 0.00859 - 0.0619u_c + 0.0255u_c^2 - (0.0429 - 0.108u_c + 0.0120u_c^2)\eta_{100}, \label{eq:17}
\end{gather}
where $u_c$ is the central density of the neutron star normalized by the saturation density and $\eta_{100}$ is the value of $\eta$ normalized by 100 MeV \cite{SIOO14}. We note that these empirical formulae are valid in the range of $0.9\lesssim u_c \le 2.0$. Since $z$ is a combination of $M$ and $R$, one can plot the mass and radius relation once one selects the value of $\eta$.

\begin{figure}[t]
\begin{center}
\includegraphics[width=10 cm]{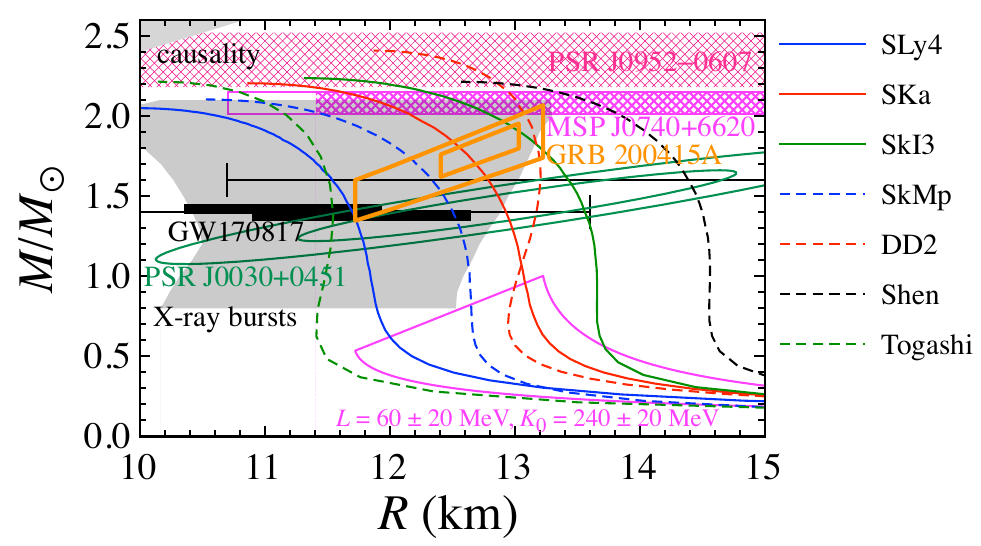}
\end{center}
\caption{Mass and radius of GRB 200415A constrained by identifying the observed high-frequency QPOs with the overtones of crustal torsional oscillations (double-parallelogram), adopting the fiducial values of $L$ and $K_0$ (or the optimal value of $L$ to identify the magnetar QPOs). For reference, the other constraints obtained from the astronomical observations and experimental restrictions are also shown. For astronomical observations, as well as those shown in Fig. \ref{fig1}, we plot the constraints of the $1.6M_\odot$ neutron star radius, i.e., $R_{1.6}\gtrsim 10.7$ km \cite{BJJS17}, the $1.4M_\odot$ neutron star radius, i.e., $R_{1.4}=11.0^{+0.9}_{-0.6}$ km~\cite{Capano20} or $R_{1.4}=11.75^{+0.86}_{-0.81}$ km \cite{Dietrich20} obtained from the GW170817; and the massive neutron star mass of PSR J0952-0607, i.e., $M=2.35\pm 0.17M_\odot$ \cite{Romani22}. Additionally, the mass and radius relations for neutron stars constructed with several EOSs as in Fig.~\ref{fig1} are also plotted.
Taken from \cite{SKS23}.\label{fig19}}
\end{figure}   

The values of $\varsigma$ and $\varsigma_{QPO}$, which are adopted to make a constraint on the mass and radius of the neutron star shown in the left panel of Fig.~\ref{fig18}, are respectively given by $L=60\pm 20$ MeV and $L=58-73$ with $K_0=240\pm 20$ MeV, which correspond to the value of $\eta$ as $\eta=70.6-118.5$ MeV and $\eta_{QPO}=90.5-111.5$ MeV, respectively. Using this range of $\eta$, one can estimate the stellar mass and radius, whose central density is less than twice the saturation density, as the right-bottom regions in the right panel of Fig.~\ref{fig18}. Furthermore, assuming that the radius of the neutron star whose central density is larger than twice the saturation density would be almost constant as shown with the dashed lines corresponding to $R=11.73$, 12.41, 13.03, and 13.23 km for $\eta = 70.6$, 90.5, 111.5, and 118.5 MeV,  we can get the overlap region with the constraint on the neutron star mass and radius to identify the QPOs with the overtones of torsional oscillations shown in the left panel of Fig.~\ref{fig18}. In such a way, we can get the neutron star mass and radius constraint for GRB 200415A. The resultant mass and radius constraint is shown in Fig.~\ref{fig19} with a double-parallelogram, together with the other constraints obtained from the astronomical observations and experimental constraints.

\begin{figure}[t]
\begin{center}
\includegraphics[width=12 cm]{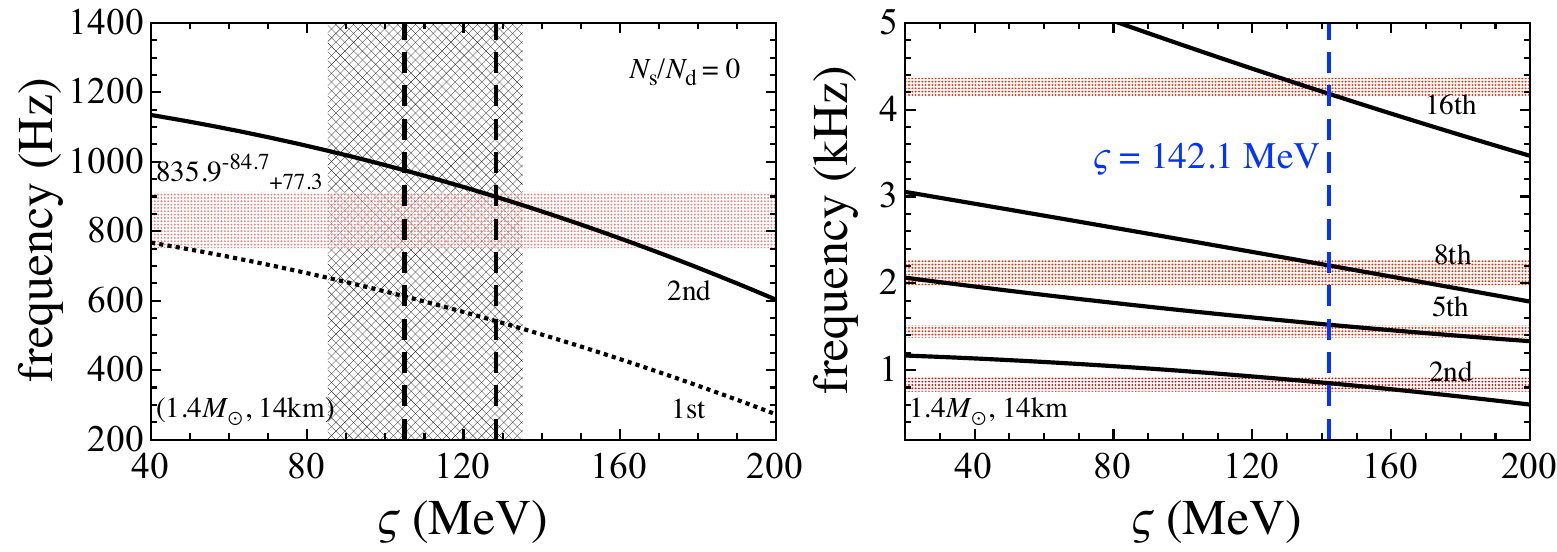}
\end{center}
\caption{ 
An alternative possibility to identify the observed QPOs in GRB 200415A with the overtones of torsional oscillations, i.e., the lowest QPO frequency is identified with the 2nd overtone instead of the 1st overtone. In the left panel, the lowest QPO frequency (shaded horizontal region) is compared with the 1st (dotted line) and 2nd (solid line) overtones as a function of $\varsigma$ for a neutron star model with $1.4M_\odot$ and 12 km. The vertical shaded region corresponds to $\varsigma=85.3-135.1$ MeV with fiducial values of $L$ and $K_0$, while the region between the vertical dashed lines is $\varsigma_{QPO}=104.9-128.4$ MeV. In the right panel, the four observed QPOs are identified with the 2nd, 5th, 8th, and 16th overtones for a neutron star model with $1.4M_\odot$ and 14 km. The vertical dashed line denotes the optimal value of $\varsigma$ with this correspondence. 
Taken from \cite{SKS23}.\label{fig20}}
\end{figure}   

Finally, we comment on the alternative identification of the QPOs observed in GRB 200415A with the overtones of torsional oscillations. To obtain the neutron star mass and radius constraint shown in Fig.~\ref{fig19}, we identify the lowest QPO observed in GRB 200415A, i.e., 835.9 Hz QPO, with the 1st overtone of the torsional oscillations, but it may be possible to identify it with the 2nd overtone. In the left panel of Fig.~\ref{fig20}, we compare the lowest QPO observed in GRB 200415A with the 1st (dotted line) and 2nd (solid line) overtones for a neutron star model with $1.4M_\odot$ and 14 km. In this figure, we also show the range of $\varsigma$ with the fiducial value of $L$ and $K_0$ (shaded region) and that with the optimal value of $L$ to identify the QPOs observed in the giant flares (the region between the vertical dashed lines). From this figure, one can find that the lowest QPO observed in GRB 200415A cannot be identified with the 2nd overtone at least with this stellar model, consistently with the region of $\varsigma$ estimated with the fiducial value of saturation parameters. In practice, one can identify the four QPOs observed in GRB 200415A with the 2nd, 5th, 8th, and 16th overtones, if $\varsigma \simeq 142$ MeV, as shown in the right panel of Fig.~\ref{fig20}. Still, anyway, it is inconsistent with the range of $\varsigma$ with the fiducial value of saturation parameters.
We note that the stellar model considered here is a little larger than the constraint from GW170817, i.e., the $1.4M_\odot$ neutron star radius should be less than 13.6 km. But, as shown in the right panel of Fig.~\ref{fig17}, the optimal value of $\varsigma$ to identify the observed QPO frequencies becomes larger as the stellar radius decreases with fixed mass. That is, if one considers the $1.4M_\odot$ neutron star model with a radius smaller than 13.6 km, the optimal value of $\varsigma$ is more distant from the estimation of $\varsigma$ using the fiducial value of $K_0$ and $L$.

\section{Shear and interface oscillations}
\label{sec:6}

Owing to the presence of the crust elasticity, the shear ($s$-) and interface ($i$-) modes are also excited as well as the torsional oscillations. To see the behavior of the $s$- and $i$-mode frequencies, again we consider the stellar model with the pasta phases, where the elasticity in the phase composed of slablike nuclei is assumed to be zero. In the same as in the torsional oscillations, one has to consider the effect of the neutron superfluidity, i.e., $N_s/N_d$ to estimate the value of $N_s$ in Eq.~(\ref{eq:7}) and ${\cal R}$ in Eq.~(\ref{eq:8}), but for simplicity we consider only the situation of the maximum enthalpy here, i.e., all the dripped neutron comove with the protons. 

In this review, we simply adopt the Cowling approximation to examine the $s$- and $i$-mode oscillations. Therefore, the perturbation equations for the $s$- and $i$-modes are derived from the linearized energy-momentum conservation laws, which are completely the same as those for the $f$-, $p_i$-, and $g_i$-modes if one neglects the elasticity. One can find the concrete perturbation equations in \cite{YL02,Sotani23}, even though they are not shown explicitly here. We note that the perturbation equations without the Cowling approximation are also found in \cite{KHA15}, which can be obtained from the linearized Einstein equations. 
We note that one can qualitatively discuss the behavior of the frequencies determined with the Cowling approximation, but the quantitative discussion may be problematic at least the $f$- and $p$-modes, which deviate $\sim 20\%$ from those determined without the Cowling approximation~\cite{YK97}. On the other hand, since the damping rate for the $s$- and $i$-modes is too small to calculate numerically, the $s$- and $i$-modes might be determined relatively well even with the Cowling approximation. In fact, it has been shown that the $g$-modes excited with the density discontinuity, whose damping rate is too small, are accurately determined with the Cowling approximation \cite{STM01}. Anyway, one has to discuss the accuracy of the $s$- and $i$-mode frequencies with the Cowling approximation somewhere. 

As the boundary conditions, one should impose the regularity condition at the center and the condition that the Lagrangian perturbation of pressure should be zero at the stellar surface. In addition, at the interface where the elasticity discontinuously becomes zero, one has to impose the junction conditions, such as the continuity of the radial displacement, the radial and transverse tractions, and Lagrangian perturbation of pressure \cite{YL02,Finn90}. In practice, with our stellar models, one has to impose the junction conditions at the four interfaces, i.e., between the fluid core and the phase of spherical-hole nuclei, between the phase of cylindrical-hole nuclei and the phase of slablike nuclei, between the phase of slablike nuclei and the phase of cylindrical nuclei, and between the surface of the outer crust and envelope. 

\begin{figure}[t]
\begin{center}
    \includegraphics[width=12 cm]{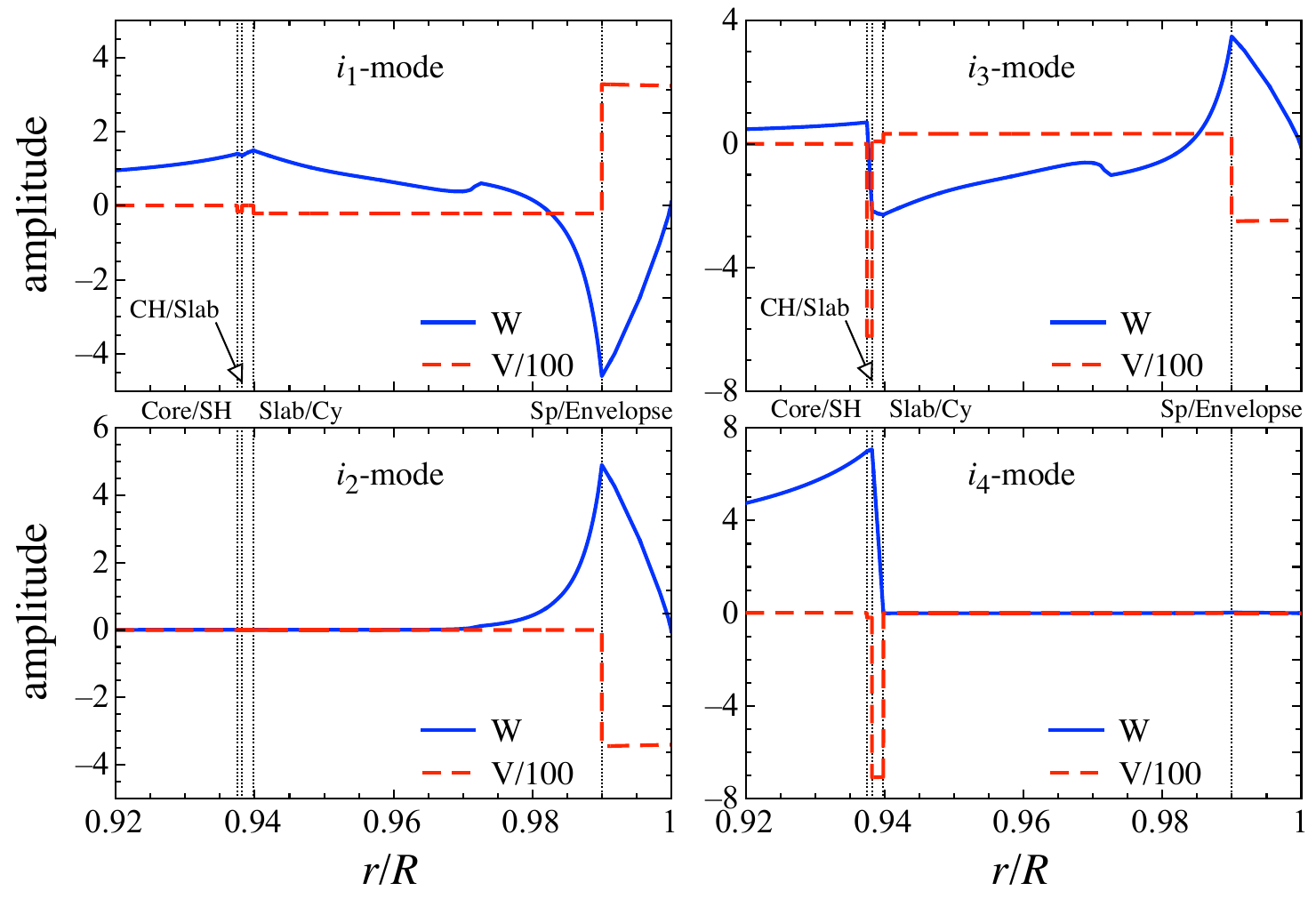}
\end{center}
\caption{ 
The radial profile of the amplitude of the Lagrangian displacement for the $i_i$-modes in the radial ($W$) and angular direction ($V$) for a stellar model with $1.44M_\odot$ and 10.2 km constructed with the EOS with $K_0=230$ MeV and $L=42.6$ MeV. The vertical dotted lines from left to right denote the interface between the fluid core and the phase of spherical-hole nuclei, the interface between the phase of cylindrical-hole and the phase of slablike nuclei, the interface between the phase of slablike nuclei and the phase of cylindrical nuclei, and the interface between the surface of crust and envelope. 
Taken from \cite{Sotani24}.\label{fig21}}
\end{figure}   

The transition density at these interfaces becomes crucial for the $s$- and $i$-mode oscillations. The transition density where the nuclear shape would be changed is determined from the EOS, depending on the saturation parameters, e.g., \cite{OI03,OI07,Oya23}. On the other hand, the transition density at the outer crust surface is more or less complicated. This is because the density to determine such properties is too low to neglect the thermal effect, even though the thermal effect on the neutron star structure is negligible when the density is very high \cite{GPE83}. In this review, we simply consider that the transition density at the surface of the outer crust is $10^{10}$ g/cm$^3$. In addition, we set the surface density, $\rho_s$, being $10^{6}$ g/cm$^3$.

\begin{figure}[t]
\begin{center}
    \includegraphics[width=6 cm]{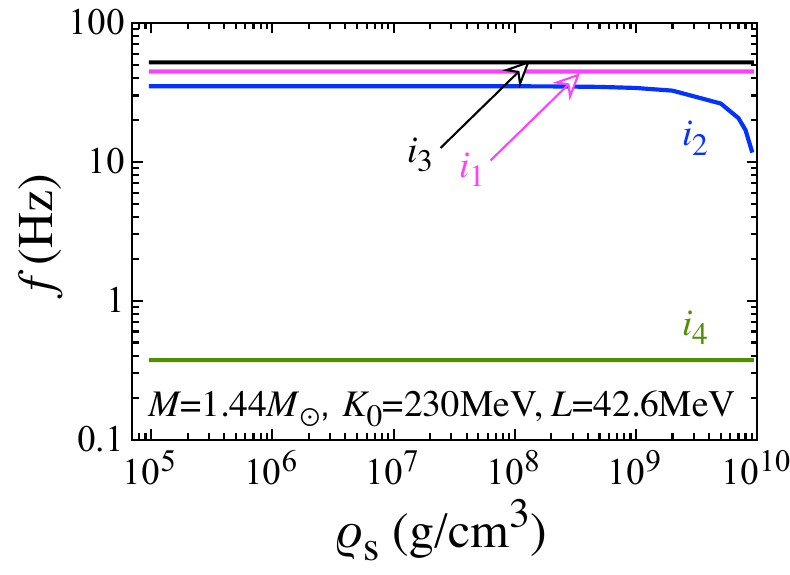}
\end{center}
\caption{ 
Dependence of the $i$-mode frequencies on the surface density, $\rho_s$, for a neutron star model with $M=1.44M_\odot$ constructed with $K_0=230$ MeV and $L=42.6$ MeV. The other transition density between the phases with and without elasticity is fixed. 
Taken from \cite{Sotani24}.\label{fig22}}
\end{figure}

In Fig.~\ref{fig21}, as an example, we show the radial profile of the amplitude of the Lagrangian displacement for the $i_i$-modes ($i=1-4$) in the radial ($W$) and angular directions ($V$) for a stellar model with $1.44M_\odot$ and 10.2 km constructed with the EOS with $K_0=230$ MeV and $L=42.6$ MeV. The classification of $i$-mode is not uniquely fixed, but here we identify that the $i_2$-mode is mainly excited at the interface between the crust surface and envelope; the $i_1$-mode is excited at the interface between the crust surface and envelope and at the interface between the phase of slablike nuclei and the phase of cylindrical nuclei; the $i_3$- and $i_4$-modes are mainly excited at the interface associated with the cylindrical-hole and spherical-hole nuclei. Since the $i$-modes are eigenmodes excited due to the existence of the interface where the elasticity discontinuously becomes zero \cite{PB05}, the number of the excited $i$-modes is generally equivalent to the number of the interface \cite{Sotani23,Sotani24}. With our classification, the order of the $i$-modes generally becomes $i_4$-, $i_2$-, $i_1$-, and $i_3$-mode from the lowest to highest, and typically the frequencies are less than $\sim 100$ Hz. Moreover, in Fig.~\ref{fig22}, we show how $i$-mode frequencies depend on the surface density, $\rho_s$, by fixing the other transition density. From this figure, one can observe the $i$-mode frequencies hardly depend on $\rho_s$. 

\begin{figure}[t]
\begin{center}
    \includegraphics[width=12 cm]{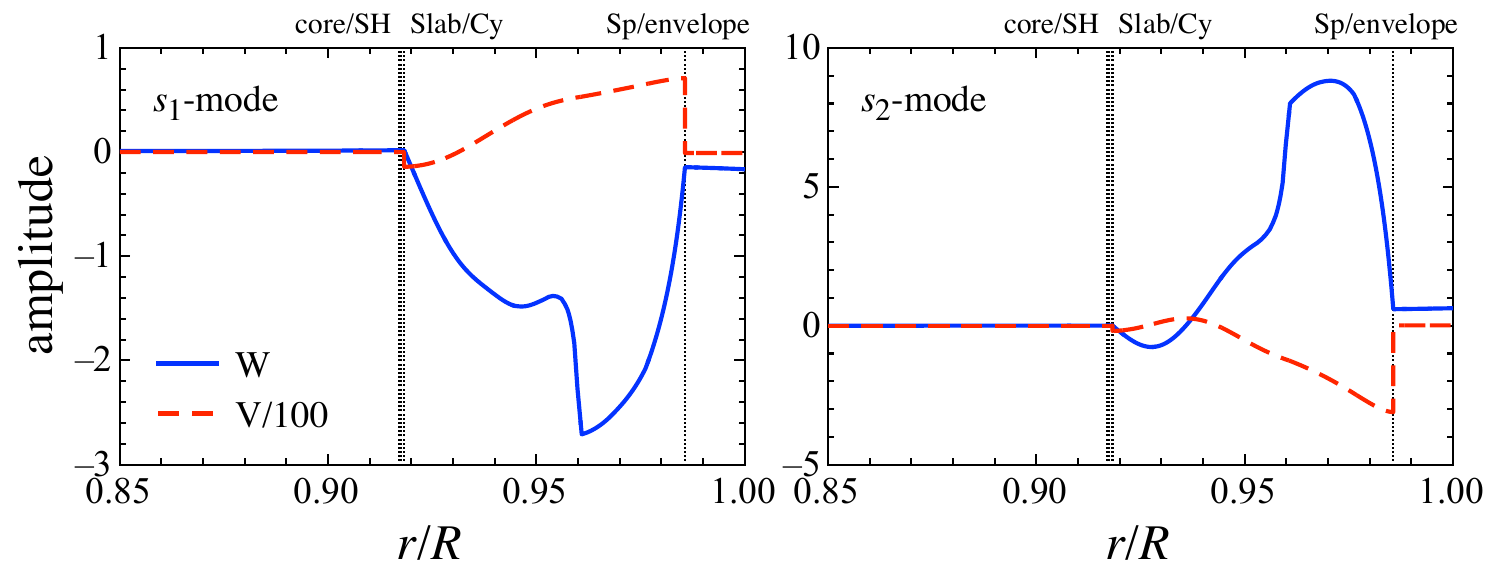}
\end{center}
\caption{ 
The radial profile of the amplitude of the Lagrangian displacement for the $s_1$- and $s_2$-modes in the radial ($W$) and angular direction ($V$) for a neutron star model with $1.4M_\odot$ and 12.4 km constructed with the EOS with $K_0=230$ MeV and $L=73.4$ MeV. 
Taken from \cite{Sotani23}.\label{fig23}}
\end{figure}   

On the other hand, in Fig.~\ref{fig23}, we show the radial profile of the amplitude of the Lagrangian displacement for the $s_1$- and $s_2$-modes in the radial ($W$) and angular directions ($V$) for a neutron star model with $1.4M_\odot$ and 12.4 km constructed with the EOS with $K_0=230$ MeV and $L=73.4$ MeV. Unlike the $i$-modes, there are an infinite number of the $s$-modes. As shown in Fig.~\ref{fig23}, the $s$-modes are excited almost only inside the elastic region. Thus, in principle, one may find the $s$-modes excited inside the phase of cylindrical-hole and spherical-hole nuclei. Even so, we could not find them in the frequency range of less than a few kHz. This situation could be understood as follows. Focusing on the angular displacement, $V$, in Fig.~\ref{fig23}, one may see that the wavelength of the $s_i$-mode, $\lambda_i$, is roughly estimated as
\begin{equation}
  \lambda_i \simeq 2\Delta R/i, 
  \label{eq:18}
\end{equation}
where $\Delta R$ denotes the thickness of the elastic region where the $s$-mode is confined. Meanwhile, the $s_i$-mode frequency may be estimated as
\begin{equation}
  f_{s_i} \simeq v_s / \lambda_i, 
  \label{eq:19}
\end{equation}
using the shear velocity, $v_s$, \cite{KHA15}. With this simple estimation, one can expect that, even if the $s$-modes are excited inside the phase of cylindrical-hole and spherical-hole nuclei, their frequencies must become much higher than those excited in the phase of spherical and cylindrical nuclei. Namely, $\Delta R$ for the phase of cylindrical-hole and spherical-hole nuclei, $\Delta R_{CHSH}$, is much thinner than $\Delta R$ for the phase of spherical and cylindrical nuclei, $\Delta R_{SpCy}$, i.e., $\Delta R_{CHSH}/\Delta R_{SpCy}\sim 1/100$, which leads to the estimation that the $s$-mode frequencies excited in the phase of cylindrical-hole and spherical-hole nuclei must be $\sim 100$ times higher than those excited in the phase of spherical and cylindrical nuclei.

Since the crust thickness, $\Delta R$, also depends on the stellar compactness~\cite{PR95,SIO17b}, to see the dependence of the $s$- and $i$-mode frequencies on the neuron star properties, we consider not only the neutron star models constructed with the original OI-EOSs but also those constructed with the EOS, which is constructed in such a way that the OI-EOS for a lower density region ($\varepsilon\le \varepsilon_t$) is connected to a one-parameter EOS given by
\begin{equation}
  p = \alpha(\varepsilon - \varepsilon_t) + p_t, 
  \label{eq:20}
\end{equation}
for a higher density region ($\varepsilon\ge \varepsilon_t$), where we especially adopt that $\varepsilon_t$ is equivalent to twice the saturation density \cite{Sotani17}. In the one parameter EOS, $p_t$ is given from the OI-EOS at $\varepsilon=\varepsilon_t$ and $\alpha$ is related to the sound velocity, $c_s$, as $c_s^2=\alpha$. Thus, $\alpha$ is a kind of indicator for expressing the stiffness of the core (or higher density) region. In this review, we focus on $1/3\le\alpha\le 1$.

\begin{figure}[t]
\begin{center}
    \includegraphics[width=12 cm]{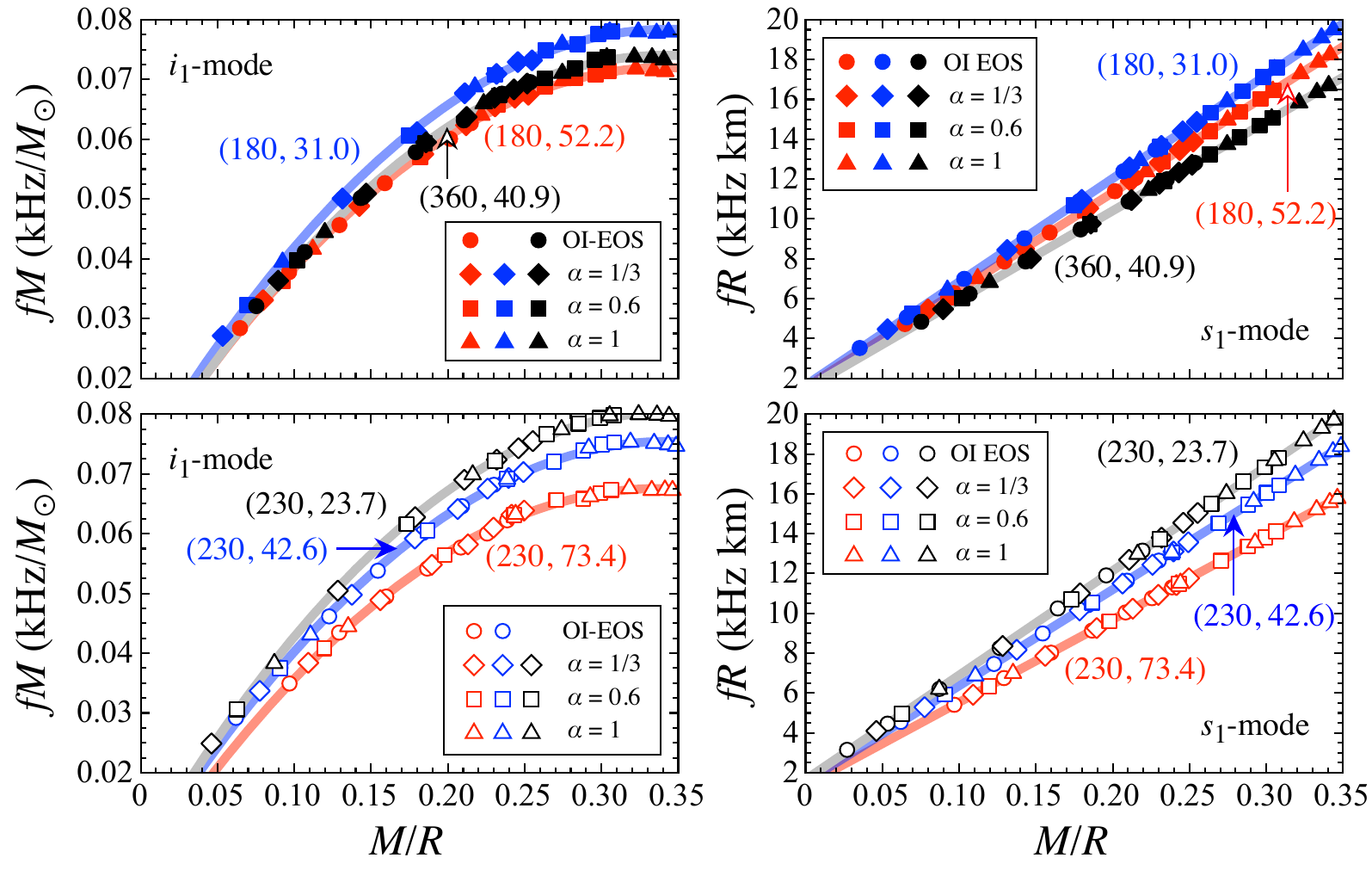}
\end{center}
\caption{ 
For various neutron star models, $f_{i_1}M$ and $f_{s_1}R$ are respectively plotted as a function of $M/R$ in the left and right panels, where the circles, diamonds, squares, and triangles denote the results for stellar models constructed with the original OI-EOS, OI-EOS connected to the EOS with $\alpha=1/3$, 0.6, and 1, respectively. The solid lines denote the fitting by Eqs.~(\ref{eq:21}) in the left panel and~(\ref{eq:22}) in the right panel, while the values of the EOS parameters $(K_0,L)$ are also shown on each line.
Taken from \cite{Sotani24}.\label{fig24}}
\end{figure}   

Through a systematical study, we find empirical relations for the $i_i$-mode frequency, $f_{i_i}$, and the $s_i$-mode frequency, $f_{s_i}$. That is, $f_{i_i}M$ and $f_{s_i}R$ can be expressed as a function of the stellar compactness, $x$, given by $x\equiv M_{1.4}/R_{12}$ with $M_{1.4}=M/1.4M_\odot$ and $R_{12}=R/12 {\rm km}$, almost independently of the stiffness of the core region, i.e., the value of $\alpha$, such as
\begin{gather}
  f_{i_i} M\ ({\rm kHz}/M_\odot) = a_{0i} + a_{1i} x + a_{2i} x^2, \label{eq:21} \\
  f_{s_i} R\ ({\rm kHz\ km}) = b_{0i} + b_{1i} x, \label{eq:22}
\end{gather}
where $a_{0i}$, $a_{1i}$, $a_{2i}$, $b_{0i}$, and $b_{1i}$ are the adjusted coefficients depending on the crust properties \cite{Sotani23}. As an example, in Fig.~\ref{fig24}, we show the results for the $i_1$-mode (left panel) and the $s_1$-mode (right panel), where the solid lines denote the fitting given by Eqs.~(\ref{eq:21}) and (\ref{eq:22}). Thus, once one simultaneously observes the shear and interface mode frequencies from a neutron star, one might extract the neutron star mass and radius via the empirical relations given by Eqs.~(\ref{eq:21}) and~(\ref{eq:22}) with the help of the constraint on the crust stiffness obtained from terrestrial experiments. Furthermore, the coefficients in Eqs.~(\ref{eq:21}) and~(\ref{eq:22}) can be well fitted as a function of a suitable combination of $K_0$ and $L$~\cite{Sotani23}. Namely, $f_{i_i} M$ and $f_{s_i}R$ can be expressed as a function of $M/R$ and a combination of $K_0$ and $L$. In practice, the $s$-mode frequency for a canonical neutron star can be estimated within $\sim 1\%$ accuracy.

\begin{figure}[t]
\begin{center}
    \includegraphics[width=12 cm]{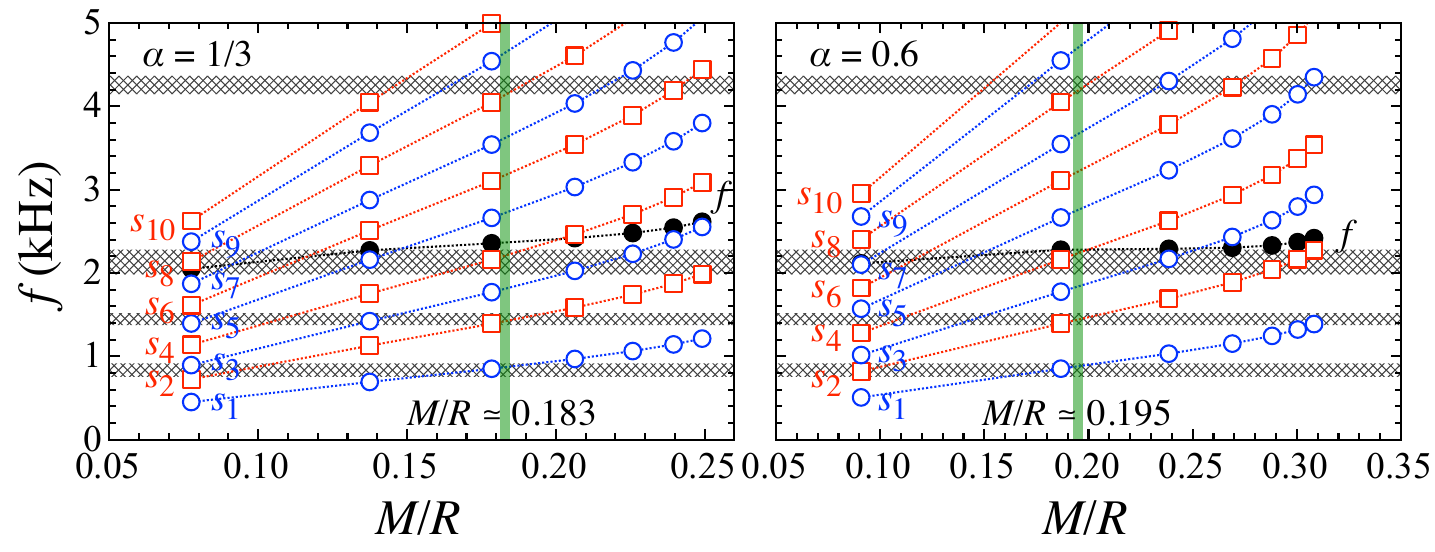}
\end{center}
\caption{ 
The $f$- and $s_i$-mode ($i=1-10$) frequencies are shown as a function of $M/R$ for neutron star models constructed with the EOS parameter of $(K_0,L)= (230,42.6)$, where the left and right panels correspond to the results for $\alpha=1/3$ and 0.6, respectively. The horizontal shaded regions denote the QPO frequencies observed in GRB 200415A \cite{C21}. The vertical lines denote the optimal value of $M/R$ to identify the four observed QPOs with the $s_i$-modes, i.e., $M/R\simeq 0.183$ for $\alpha=1/3$ and $M/R\simeq 0.195$ for $\alpha=0.6$. 
Taken from \cite{Sotani24}.\label{fig:QPO-is}}
\end{figure}   

In the end, we consider the possibility of applying the $s$-mode frequency for identifying the QPO observations. As discussed in Sec.~\ref{sec:5d}, one can identify the high-frequency QPOs observed in GRB 200415A with the overtones of crustal torsional oscillations. Since only high-frequency QPOs have been observed in this event, one may identify them with the other neutron star oscillations. In practice, as shown in Fig.~\ref{fig:QPO-is} where the $f$- and $s_i$-mode frequencies for the stellar models with $(K_0,L)=(230,42.6)$ are shown as a function of the stellar compactness, together with the QPO frequencies observed in GRB 200415A, one can identify them with the $s$-modes. Namely, the observed QPOs are identified with the $s_1$-, $s_2$-, $s_4$-, and $s_8$-mode frequencies if $M/R \simeq 0.183$ for $\alpha = 1/3$ (left panel) or $M/R \simeq 0.195$ for $\alpha = 0.6$ (right panel). One could discuss the constraint on the saturation parameters through the identification of the QPOs with the torsional oscillations in Sec.\ref{sec:5}, because the torsional oscillations are confined inside the elastic region. On the other hand, since the $s$-mode oscillations depend on not only the crust region (elastic region) but also the (fluid) core region, it may be more difficult to discuss the crust properties via the identification of QPOs with the $s$-modes. Anyway, to get a severe constraint on the neutron star properties or to validate a theoretical model, we have to wait for the next signals from a neutron star.

\section{Discussion and Conclusions}
\label{sec:7}


In this review, we discussed the neutron star oscillations excited by the crust elasticity, such as the torsional, interface, and shear oscillations, adopting the Cowling approximation. Since the torsional oscillations are axial-type oscillations, one can fully discuss them even with the Cowling approximation. On the other hand, since the interface and shear oscillations are polar-type oscillations, one can qualitatively discuss them with the Cowling approximation but one may have to check the accuracy of the interface and shear mode frequencies by comparing them to those taken into account metric perturbations. In any case, through the identification of the QPOs observed in the magnetar flares with the torsional oscillations, we successfully obtain the constraint on the nuclear saturation parameters, especially the density dependence of the nuclear symmetry energy, $L$, as $L\simeq 58-73$ MeV. On the other hand, via systematical study of the interface and shear oscillations, we find empirical relations almost independent of the stiffness of the core region. 

To determine these frequencies, the effect of the superfluidity is important input physics as well as the shear modulus. To deal with the superfluidity, in this review we introduce the parameter, such as  $N_s/N_d$ or ${\cal R}$, which is still quite uncertain. In addition, the understanding of the shear modulus inside the pasta phases may still be poor, where one has room for improvement. One may have to improve these properties to consider a more realistic situation.

As shown in this review, the observed QPO frequencies can be identified with the crustal torsional oscillations, but still one has to solve the time-dependence of the QPO excitations.  For this purpose, one may have to examine the crustal oscillations beyond linear analysis, e.g., taking into account the nonlinear coupling, and discuss the energy transfer from specific modes to other modes. 

Moreover, in this review, we consider only the oscillations excited due to the crust elasticity. Meanwhile, the existence of quark matter inside the neutron star is theoretically suggested. In such a situation, the hadronic matter would transition into quark matter, and the mixed phase may appear at the transition region, where the structure becomes similar to the pasta structure in the crust \cite{{Maruyama07,Xia20}}. So, such a mixed phase behaves as a liquid crystal, i.e., the elasticity becomes non-zero. That is, if such a phase exists, one can expect that the torsional oscillations and/or shear and interface oscillations are excited in the hadron-quark mixed phase \cite{SMT13,LY21}. Through the observations of these oscillation modes, one may probe the properties of quark matter.



\section{Patents}

\funding{This research was funded by Japan Society for the Promotion of Science (JSPS) KAKENHI Grant Numbers JP19KK0354 and JP21H01088, and by FY2023 RIKEN Incentive Research Project.
}

\institutionalreview{Not applicable.
}

\informedconsent{Not applicable.
}

\dataavailability{Dataset available on request from the authors.
} 



\acknowledgments{We are grateful to Kazuhiro Oyamatus for providing the EOS data, which is crucial for calculations in this study, and also to Kei Iida, Ken'ichiro Nakazato, Kostas D. Kokkotas, and Nick Stergioulas for valuable comments. 
}

\conflictsofinterest{The authors declare no conflicts of interest. The funders had no role in the design of the study; in the collection, analyses, or interpretation of data; in the writing of the manuscript; or in the decision to publish the results.
} 



\abbreviations{Abbreviations}{
The following abbreviations are used in this manuscript:\\

\noindent 
\begin{tabular}{@{}ll}
QPOs & Quasi-periodic oscillations\\
SGR & Soft-gamma repeater\\
EOSs & Equations of state\\
TOV & Tolman-Oppenheimer-Volkoff
\end{tabular}
}




\begin{adjustwidth}{-\extralength}{0cm}

\reftitle{References}

\PublishersNote{}
\end{adjustwidth}
\end{document}